\documentclass[twocolumn,aps,showpacs,prb,tightenlines,amsmath,amssymb,superscriptaddress]{revtex4}
\usepackage{graphicx}
\usepackage{amssymb}
\usepackage{amsmath}
\usepackage{colordvi}

\begin{document}

\title{Electron spin relaxation in paramagnetic Ga(Mn)As quantum wells}

\author{J. H. Jiang}
\affiliation{Hefei National Laboratory for Physical Sciences at
  Microscale and Department of Physics, University of Science and
  Technology of China, Hefei, Anhui, 230026, China}
\author{Y. Zhou}
\affiliation{Hefei National Laboratory for Physical Sciences at
  Microscale and Department of Physics, University of Science and
  Technology of China, Hefei, Anhui, 230026, China}
\author{T. Korn}
\affiliation{Institut f\"ur Experimentelle und Angewandte Physik,
  Universit\"at Regensburg, D-93040 Regensburg, Germany}
\author{C. Sch\"uller}
\affiliation{Institut f\"ur Experimentelle und Angewandte Physik,
  Universit\"at Regensburg, D-93040 Regensburg, Germany}
\author{M. W. Wu}
\thanks{Author to whom correspondence should be addressed}
\email{mwwu@ustc.edu.cn.}
\affiliation{Hefei National Laboratory for Physical Sciences at
  Microscale and Department of Physics, University of Science and
  Technology of China, Hefei, Anhui, 230026, China}
\date{\today}

\begin{abstract}

Electron spin relaxation in paramagnetic Ga(Mn)As quantum wells is
studied via the fully microscopic kinetic spin Bloch equation
approach where all the scatterings, such as the electron-impurity,
electron-phonon, electron-electron Coulomb, electron-hole Coulomb,
electron-hole exchange (the Bir-Aronov-Pikus mechanism) and the
$s$-$d$ exchange scatterings, are explicitly included. The
Elliot-Yafet mechanism is also incorporated. From this approach, we
study the spin relaxation in both $n$-type and $p$-type Ga(Mn)As
quantum wells. For $n$-type Ga(Mn)As quantum wells where most Mn
ions take the interstitial positions, we find that the spin
relaxation is always dominated by the DP mechanism in the
metallic region. Interestingly, the Mn concentration dependence of
the spin relaxation time is nonmonotonic and exhibits a peak.
This is due to the fact that the momentum
scattering and the inhomogeneous broadening have different density
dependences in the non-degenerate and degenerate regimes. For
$p$-type Ga(Mn)As quantum wells, we find that the Mn
concentration dependence of the spin relaxation time 
is also nonmonotonic and shows a
peak. Differently, the
cause of this behaviour is that the $s$-$d$ exchange scattering (or
the Bir-Aronov-Pikus) mechanism dominates the spin relaxation in the
high Mn concentration regime at low (or high) temperature, whereas
the DP mechanism determines the spin relaxation in the low Mn
concentration regime. The Elliot-Yafet mechanism also contributes
to the spin relaxation
 at intermediate temperatures. The spin relaxation
time due to the DP mechanism increases with increasing Mn concentration due to
motional narrowing, whereas those due to the spin-flip mechanisms
decrease with it, which thus leads to the formation of
the peak. The temperature, photo-excitation density and magnetic
field dependences of the spin relaxation time 
in $p$-type Ga(Mn)As quantum wells are
investigated systematically with the underlying physics revealed.
Our results are consistent with the recent experimental findings.

\end{abstract}


\maketitle

\section{Introduction}

Semiconductors doped with magnetic impurities have intrigued much
interest since the invention of ferromagnetic III-V semiconductors due
to the possibility of integrating both the magnetic (spin) and charge
degree of freedom on one
chip.\cite{Munekata1,Ohno1,Munekata2,Ohno2,Hayashi,VanEsch,Ohno3} Many
new device conceptions and functionalities based on these materials
are proposed, and the material properties together with the underlying
physics are extensively
studied.\cite{FStheory,MSreview,Wolf,spintronics,Dietl,Ohno2,Fabian,Back,Wagner,Saha}
Specifically, ferromagnetic Ga$_{1-x}$Mn$_x$As has been used as a
highly efficient source to inject spin polarization into
GaAs\cite{Ciorga} and magnetic tunneling junctions based on
ferromagnetic Ga$_{1-x}$Mn$_x$As can achieve very high
magnetoresistance.\cite{Ruster} Besides, the ability to detect the
magnetic moment via Hall measurements\cite{Ohno2,Tang} and to control
it via gate-voltage\cite{Ohno4} and laser radiation\cite{opt-mag}
opens the way for incorporating opto-electronics with
magnetism. Magneto-optical measurements, which could characterize the
spin splitting of carriers due to both the external magnetic field and
the $s$-$d$ or $p$-$d$ exchange field, provide important information
about the microscopic properties of the carriers, such as, the
$g$-factor, the $s$-$d$ and $p$-$d$ exchange coupling constants, as
well as the electron spin relaxation time (SRT). Such measurements
have recently been performed in dilutely-doped paramagnetic Ga(Mn)As
quantum wells.\cite{Awsch,Schulz,Korn,MOreview} Although many properties of
Ga(Mn)As have been extensively studied, the electron spin relaxation
has not yet been well understood even in dilutely doped paramagnetic
phase.  This is the aim of this investigation. We focus on (001)
Ga(Mn)As quantum wells.

Electron spin relaxation in non-magnetic GaAs has been extensively
studied and three main spin relaxation mechanisms have been
recorgnized for decades:\cite{opt-or} the D'yakonov-Perel' (DP)
mechanism,\cite{DP} the Bir-Aronov-Pikus (BAP) mechanism\cite{BAP}
and the Elliot-Yafet (EY) mechanism.\cite{EY} Usually, the DP
mechanism dominates the spin relaxation in $n$-type quantum
wells.\cite{spintronics,lowT} The BAP mechanism was believed to be
most important at low temperature in intrinsic and $p$-type quantum
wells for a long time.\cite{opt-or,spintronics} Recently, Zhou and
Wu showed that the BAP mechanism was exaggerated in the low
temperature regime in previous treatments based on the elastic
scattering approximation, where the Pauli blocking was not
considered.\cite{wu-bap} It was then found that the BAP mechanism is less
efficient than the DP mechanism in intrinsic quantum wells and
$p$-type quantum wells with high photo-excitation density. A
similar conclusion was also obtained in bulk GaAs very
recently.\cite{bulk} Previously, the EY mechanism was believed to
dominate the spin relaxation in heavily doped samples at low
temperature. However, it was shown to be unimportant in bulk GaAs by
our recent investigation.\cite{bulk} Whether this is still true in
quantum-well systems remains unchecked. Moreover, in
paramagnetic Ga(Mn)As quantum wells, things are more complicated:
(i) All the three mechanisms could be important as the material is
heavily-doped with Mn and the hole density is generally very
high;\cite{Awsch} (ii) An additional spin relaxation mechanism due
to the exchange coupling of the electrons and the localized Mn spins
(the $s$-$d$ exchange scattering mechanism) may also be
important.\cite{FStheory} In this work, we will compare different
spin relaxation mechanisms for various Mn concentrations,
temperatures, photo-excitation densities and magnetic fields.

In Ga(Mn)As, the Mn dopants can be either substitutional or
interstitial: the substitutional Mn accepts one electron, whereas
the interstitial Mn releases two. Direct doping in
low-temperature molecular-beam epitaxy growth gives rise to
more substitutional Mn ions than  interstitial
ones, which makes the Ga(Mn)As a $p$-type
semiconductor.\cite{FStheory,MSreview,Awsch} Recently, it was found
that in GaAs quantum wells near a Ga(Mn)As layer, the Mn dopants can
diffuse into the GaAs quantum well, where the Mn ions mainly take
the interstitial positions, making the quantum well
$n$-type.\cite{Edmonds,Schulz,Korn} The experimental results also
show interesting features of the SRT.

We apply the fully microscopic kinetic spin Bloch equation (KSBE)
approach\cite{wu-early,hot-e,highP} to investigate the spin relaxation
in paramagnetic Ga(Mn)As quantum wells. The KSBE approach explicitly includes all
relevant scatterings, such as, the electron-impurity, electron-phonon,
electron-electron Coulomb, electron-hole Coulomb,
electron-hole exchange (the BAP mechanism) and $s$-$d$ exchange
scatterings. Previously, the KSBE approach has been applied to study
the spin dynamics in semiconductor and its nanostructures where good
agreements with experiments have been achieved and many predictions
have been confirmed by
experiments.\cite{wu-early,wu-strain,wu-hole,highP,hot-e,lowT,wu-bap,terahertz,wu-exp,wu-exp-hP,multi-valley,multi-band,bulk,Lai,Awsch-exp,Ji,Zheng,Zheng2,tobias}
In this work, we apply the KSBE approach to both $n$- and $p$-type paramagnetic
Ga(Mn)As quantum wells to study the electron spin relaxation. We
distinguish the dominant spin relaxation mechanisms in different
regimes and our results are consistent with the recent experimental findings.

This paper is organized as follows: In Sec.~II, we set up the model
and establish the KSBEs. In Sec.~III we present our results and
discussions. We conclude in Sec.~IV.

\section{Model and KSBEs}

We start our investigation from a paramagnetic [001] grown Ga(Mn)As quantum well
of width $a$ in the growth direction (the $z$-axis). A moderate
magnetic field ${\bf B}$ is applied along the $x$-axis (the Voigt
configuration). It is assumed that the well width is small enough so
that only the lowest subband of electron and the lowest two subbands of
heavy-hole are relevant for the electron and hole densities in our
investigation. The barrier layer is chosen to be Al$_{0.4}$Ga$_{0.6}$As
where the barrier heights of electron and hole are 328 and 177~meV
respectively.\cite{Madelung} The envelope functions of the relevant
subbands are calculated via the finite-well-depth
model.\cite{lowT,wu-bap}

The KSBEs can be constructed via the nonequilibrium Green function
method\cite{Haug} and read
\begin{equation}
  \partial_t \hat{\rho}_{\bf k} =
  \left. \partial_t \hat{\rho}_{\bf k}\right|_{\rm coh}
  + \left. \partial_t \hat{\rho}_{\bf k}\right|_{\rm scat},
\end{equation}
with $\hat{\rho}_{\bf k}$ representing the single-particle density
matrix whose diagonal and off-diagonal elements describe the
electron distribution functions and the spin coherence
respectively.\cite{wu-early} The coherent term is given by
($\hbar\equiv 1$ throughout this paper)
\begin{eqnarray}
&& \hspace{-0.8cm} \left. \partial_t \hat{\rho}_{\bf k}\right|_{\rm coh} =
  \nonumber \\
  &&\hspace{-0.4cm} \mbox{} - i \left[
  \left(g_e \mu_{\rm B} {\bf B} + {\bf h}({\bf
  k})\right)\cdot\frac{\hat{\mbox{\boldmath$\sigma$\unboldmath}}}{2} +
  \hat{H}^{\rm mf}_{\rm sd} +
  \hat{\Sigma}_{\rm HF} ({\bf k}),  \hat{\rho}_{\bf k}\right] ,
\end{eqnarray}
in which $[\ ,\ ]$ is the commutator. $g_e$ is the electron
$g$-factor. ${\bf h}({\bf k})$ represents the spin-orbit coupling
(SOC), which is composed of the Dresselhaus\cite{Dresselhaus}
and Rashba\cite{Rashba} terms. In symmetric GaAs quantum well with
small well width, the Dresselhaus term is dominant\cite{Flatte} and
\begin{eqnarray}
  {\bf h}({\bf k})=\gamma_{\rm D}\left(k_x \left(k_y^2-\langle k_z^2 \rangle \right),\
  k_y\left( \langle k_z^2 \rangle-k_x^2 \right),\ 0\right).
\end{eqnarray}
Here $\langle k_z^2 \rangle$ represents the average of the operator
$-(\partial / \partial z)^2$ over the state of the lowest electron
subband and $\gamma_{\rm D}=11.4$~eV$\cdot$\AA$^3$ is the Dresselhaus SOC
coefficient.\cite{lowT} The mean-field contribution of the $s$-$d$
exchange interaction is given by
\begin{equation}
  \hat{H}^{\rm mf}_{\rm sd} = -N_{\rm Mn}{\alpha} \langle {\bf S} \rangle
    \cdot \frac{\hat{\mbox{\boldmath$\sigma$\unboldmath}}}{2},
    \label{equ:H_mf_ex}
\end{equation}
where $\langle {\bf S} \rangle$ is the average spin polarization of Mn
ions and $\alpha$ is the $s$-$d$ exchange coupling constant. For simplicity,
we assume that the Mn ions are uniformly distributed within and around
the Ga(Mn)As quantum well with a bulk density $N_{\rm Mn}$.
$\hat{\Sigma}_{\rm HF} ({\bf k})=-\sum_{{\bf q},q_z}
V_{{\bf q},q_z}|I(iq_z)|^2\hat{\rho}_{{\bf k}-{\bf q}}$ is the Coulomb
Hartree-Fock (HF) term, where $I(iq_z)=\int dz |\xi_e(z)|^2e^{iq_z z}$
is the form factor with $\xi_e(z)$ standing for the envelope
function of the lowest electron subband.\cite{highP} $V_{{\bf q},q_z}$
is the screened Coulomb potential. In this work, we take into account
the screening from both electrons and holes within the random phase
approximation.\cite{wu-bap}

The scattering term $\left. \partial_t \hat{\rho}_{\bf
  k}\right|_{\rm scat}$ consists of the electron-impurity,
electron-electron Coulomb, electron-phonon, electron-hole Coulomb,
electron-hole exchange and $s$-$d$ exchange scatterings. The
expressions of all these terms except the $s$-$d$ exchange scattering
can be found in Ref.~\onlinecite{wu-bap}. However, the expression of
the electron-impurity scattering term with the EY mechanism
included has not been given in that paper, which we will present
later in this paper. The $s$-$d$ exchange scattering term is given by
\begin{eqnarray}
  \left. \partial_t \hat{\rho}_{\bf k}\right|_{\rm sd}^{\rm scat} &=&
  -\pi N_{\rm Mn} \alpha^2 I_s \mspace{-10mu}\sum_{\eta_1 \eta_2 {\bf
      k}^{\prime}} \mspace{-10mu} G_{\rm Mn}(-\eta_1{-\eta_2})
  \delta(\varepsilon_{\bf k}-\varepsilon_{{\bf k}^{\prime}})
    \nonumber\\ && \mbox{} \times
  \Big[ \hat{s}^{\eta_1} \hat{\rho}^>_{{\bf k}^{\prime}}
    \hat{s}^{\eta_2} \hat{\rho}^<_{\bf k}
    - \hat{s}^{\eta_2}
    \hat{\rho}^<_{{\bf k}^{\prime}}
    \hat{s}^{\eta_1}
    \hat{\rho}^>_{\bf k} + {\rm H.c.}
    \Big].
  \label{equ:rho_ex_scat0}
\end{eqnarray}
Here $\hat{\rho}^>_{\bf k}=\hat{1}-\hat{\rho}_{\bf k}$,
$\hat{\rho}^<_{\bf k}=\hat{\rho}_{\bf k}$, $G_{\rm
  Mn}(\eta_1\eta_2)=\frac{1}{4}{\rm
  Tr}(\hat{S}^{\eta_1}\hat{S}^{\eta_2}\hat{\rho}_{\rm Mn})$
and $I_s=\int dz |\xi_e(z)|^4$.
$\hat{S}^{\eta}$ and $\hat{s}^{\eta}$ ($\eta=0,\pm 1$) are the spin ladder
operators with $\hat{S}^0=\hat{S}_z$, $\hat{S}^{\pm}=\hat{S}_x\pm i
\hat{S}_y$, $\hat{s}^0=2\hat{s}_z$ and $\hat{s}^{\pm}=\hat{s}_x\pm i
\hat{s}_y$. $\hat{\rho}_{\rm Mn}$ is the Mn spin density matrix.
$\varepsilon_{\bf k}=k^2/2m^{\ast}$ is the electron kinetic energy
  with $m^{\ast}$ denoting the effective mass.
The equation-of-motion for Mn spin density matrix consists of three parts
$ \partial_t \hat{\rho}_{\mathrm{Mn}} = \left. \partial_t
\hat{\rho}_{\mathrm{Mn}}\right|_{\rm coh}  + \left. \partial_t
\hat{\rho}_{\mathrm{Mn}}\right|_{\rm scat} + \left. \partial_t
\hat{\rho}_{\mathrm{Mn}}\right|_{\rm rel}$. The first part describes the
coherent precession around  the external mangetic field and the
$s$-$d$ exchange mean field, $\left. \partial_t
\hat{\rho}_{\mathrm{Mn}}\right|_{\rm coh} = - i \left[ g_{\rm Mn} \mu_{\rm B}
  {\bf B}\cdot \hat{{\bf S}}  - \alpha\sum_{{\bf k}}{\rm
    Tr} (\frac{\hat{\mbox{\boldmath$\sigma$\unboldmath}}}{2} \hat{\rho}_{\bf
    k}) \cdot \hat{{\bf S}},\ \hat{\rho}_{\mathrm{Mn}}\right]$.
The second part represents the $s$-$d$ exchange scattering with
electrons $\left. \partial_t \hat{\rho}_{\mathrm{Mn}}\right|_{\rm scat}
= -\frac{\pi \alpha^2}{4} \sum_{\eta_1 \eta_2 {\bf k}} 
\delta(\varepsilon_{\bf k}-\varepsilon_{{\bf k}^{\prime}})
{\rm Tr}(\hat{s}^{-\eta_2} \hat{\rho}^<_{{\bf k}^{\prime}}
\hat{s}^{-\eta_1} \hat{\rho}^>_{\bf k} )
\Big[ (\hat{S}^{\eta_1}\hat{S}^{\eta_2} \hat{\rho}_{\mathrm{Mn}}
  - \hat{S}^{\eta_1}\hat{\rho}_{\mathrm{Mn}}\hat{S}^{\eta_2}) 
  + {\rm H.c.}  \Big]$. The third term characterizes the Mn spin
relaxation due to other mechanisms, such as, $p$-$d$ exchange
interaction with holes or Mn-spin--lattice interactions, with a
relaxation time approximation,
$ \left. \partial_t \hat{\rho}_{\mathrm{Mn}}\right|_{\rm rel} = -\left(
\hat{\rho}_{\mathrm{Mn}}-\hat{\rho}_{\mathrm{Mn}}^{0}
\right)/\tau_{\mathrm{Mn}}$. Here $\hat{\rho}_{\mathrm{Mn}}^{0}$
represents the equilibrium Mn spin density
matrix. $\tau_{\mathrm{Mn}}$ is the Mn spin relaxation time, which
is typically 0.1$\sim$10~ns.\cite{Myers} In our calculation we take
$\tau_{\mathrm{Mn}}=1$~ns. At $t=0$, the Mn spin density matrix is
chosen to be the equilibrium one $\hat{\rho}_{\mathrm{Mn}}^{0}$. 
The Mn spins can be dynamically polarized via the $s$-$d$ exchange
interaction, and feedback to the electron spin dynamics. However, 
we find that this process affects the electron spin dynamics
little. Hence, the choice of $\tau_{\mathrm{Mn}}$ does not affect our
discussions on electron spin dynamics.

The $s$-$d$ exchange scattering $\tau_{\rm sd}$ can be obtained
analytically. In the absence of an external magnetic field,
the spin polarization of the electron system is always along the
$z$-direction. As the $s$-$d$ exchange interaction conserves the
spin polarization of the total system, the spin polarization of Mn
ions (which is assumed to be zero initially) can only be along
the $z$-direction. By keeping only the diagonal term of
$\hat{\rho}_{\rm
  Mn}$, from the Fermi Golden rule, the spin relaxation time due to the
$s$-$d$ exchange scattering can be obtained directly:
\begin{eqnarray}
  \tau_{\rm sd} &=& \left\{\frac{1}{2}N_{\rm Mn}I_s\alpha^2m^\ast[S(S+1)
    -\langle S_z^2 \rangle] \right\}^{-1} \nonumber\\
  &=&\frac{12}{35N_{\rm Mn}\alpha^2m^\ast I_s} ,
  \label{tau_ex}
\end{eqnarray}
where $S=5/2$ is the spin of the Mn ion. It is evident that
$\tau_{\rm sd}$ is independent of temperature and electron density, but
is inverse proportional to Mn concentration $N_{\rm Mn}$, the square
of the $s$-$d$ exchange coupling $\alpha^2$ and $I_s=\int dz
|\xi_e(z)|^4$. $I_s$ is determined by the confinement of the quantum
well. For an infinite-depth-well $I_s=\frac{3}{2a}$, thus
$\tau_{\rm sd}$ is proportional to the well width $a$.

After incorporating the EY mechanism, besides the ordinary
spin-conserving term, there are spin-flip terms. For
electron-impurity scattering these additional terms are
\begin{eqnarray}
  \left. \partial_t \hat{\rho}_{\bf k}\right|_{\rm ei}^{\rm EY} &=& -\pi
  n_{i} \sum_{{\bf k}^{\prime}} \delta(\varepsilon_{\bf
  k}-\varepsilon_{{\bf k}^{\prime}}) \Big[ U^{(1)}_{{\bf k}-{\bf
  k}^\prime} \big( \hat{\Lambda}^{(1)}_{{\bf k},{\bf k}^\prime}
  \hat{\rho}^>_{{\bf k}^{\prime}}
  \hat{\Lambda}^{(1)}_{{\bf k}^\prime,{\bf k}}
  \nonumber \\ && \mbox{} \times
    \hat{\rho}^<_{\bf k} - \hat{\Lambda}^{(1)}_{{\bf k},{\bf k}^\prime}
  \hat{\rho}^<_{{\bf k}^{\prime}} \hat{\Lambda}^{(1)}_{{\bf k}^\prime,{\bf k}}
    \hat{\rho}^>_{\bf k} \big)
    + U^{(2)}_{{\bf k}-{\bf k}^\prime}
    \big( \hat{\Lambda}^{(2)}_{{\bf k},{\bf k}^\prime} \hat{\rho}^>_{{\bf k}^{\prime}}
    \nonumber \\ && \mbox{} \times
\hat{\Lambda}^{(2)}_{{\bf k}^\prime,{\bf k}}
      \hat{\rho}^<_{\bf k} - \hat{\Lambda}^{(2)}_{{\bf k},{\bf k}^\prime}
    \hat{\rho}^<_{{\bf k}^{\prime}} \hat{\Lambda}^{(2)}_{{\bf k}^\prime,{\bf k}}
    \hat{\rho}^>_{\bf k} \big)
     + {\rm H.c.} \Big],
  \label{equ:rho_EY_ei}
\end{eqnarray}
where $n_i = N_{\rm Mn}^{\rm S}+4N_{\rm Mn}^{\rm I} +
n_{i0}$ with $N_{\rm Mn}^{\rm S}$, $N_{\rm Mn}^{\rm I}$, $n_{i0}$ being the
densities of substitutional Mn, interstitial Mn and non-magnetic
impurities, respectively due to different charges. $U^{(1)}_{{\bf k}-{\bf
    k}^\prime} = \frac{\lambda_c^2}{4}  \sum_{q_z} V^2_{{\bf k}-{\bf
    k}^{\prime},q_z} |I(iq_z)|^2
q_z^2$ and $U^{(2)}_{{\bf k}-{\bf k}^\prime} = - \lambda_c^2
\sum_{q_z} V^2_{{\bf k}-{\bf k}^{\prime},q_z} |I(iq_z)|^2$.
Here
$\lambda_c=\frac{\eta(1-\eta/2)}{3m_cE_g(1-\eta/3)}$ with
$\eta=\frac{\Delta_{\rm SO}}{\Delta_{\rm SO} + E_g}$. $E_g$ and $\Delta_{\rm SO}$
are the band-gap and the spin-orbit splitting of the valence band,
respectively.\cite{opt-or}
The spin-flip matrices are given by
$\hat{\Lambda}^{(1)}_{{\bf k}^\prime,{\bf k}} =
[({\bf k}+{\bf k}^{\prime},0)\times
  \hat{\mbox{\boldmath${\sigma}$\unboldmath}}]_z$ and
$\hat{\Lambda}^{(2)}_{{\bf k},{\bf k}^\prime} = [({\bf k},0)\times({\bf
    k}^{\prime},0)]\cdot
\hat{\mbox{\boldmath${\sigma}$\unboldmath}}$.
It is noted that $\hat{\Lambda}^{(1)}_{{\bf
    k}^\prime,{\bf k}}$ and $\hat{\Lambda}^{(2)}_{{\bf k}^\prime,{\bf k}}$
contribute to the out-of-plane and in-plane spin relaxations,
respectively. They are generally different and therefore the spin
relaxation due to the EY mechanism in quantum wells is anisotropic.
The EY mechanism can be incorporated into other scatterings
similarly.\cite{bulk} However, we find that the EY mechanism can
 be important only when the impurity density is high,
where the electron-impurity scattering is most important. Therefore,
for simplicity, we include only the EY spin-flip processes
associated with the electron-impurity scattering.

\section{Results and discussions}

By solving the KSBEs numerically, we obtain the temporal evolution of
the single particle density matrix $\hat{\rho}_{\bf k}$ and then
the spin polarization along the $z$-axis, i.e., $s_z$. From the decay of
$s_z$, the SRT is extracted. The initial spin polarization is chosen
to be $P=4~\%$. The well width $a=10$~nm. The external magnetic field
${\bf B}$ is zero unless otherwise specified. We use $x$ to denote the
Mn density, where $N_{\rm Mn} = x N_0$ with $N_0=\Omega^{-1}=2.22\times
10^{22}$~cm$^{-3}$ ($\Omega$ is the volume of the unit cell
in GaAs). The other material parameters used are listed in
Table~\ref{parameter}.\cite{para,Ekardt,Maialle}

The value of the $s$-$d$ exchange coupling in III(Mn)V materials is
still in dispute. In bulk Ga(Mn)As, first-principle calculation gives
the value $N_0\alpha\approx0.25$~eV.\cite{Sanvito} However, the
experimental measurements in Ref.~\onlinecite{Awsch} show that
$N_0\alpha$ is in the range of $[-0.21, -0.07]$~eV varying with
quantum well width. In this paper, we choose $N_0\alpha=-0.15$~eV
unless otherwise specified.

\begin{table}[h]
\caption{Material parameters used in the calculation}
\begin{center}
\begin{tabular}{llll}
  \hline \hline
  $\kappa_0$ & 12.9 &  $\kappa_{\infty}$ & 10.8\\
  $D$ & $5.31\times 10^3$~kg/m$^3$ & $e_{14}$ & $1.41\times 10^9$~V/m\\
  $v_{st}$ & $2.48\times 10^3$~m/s & $v_{sl}$ & $5.29\times 10^3$~m/s \\
  $\Xi$ & 8.5~eV & $\omega_{\rm LO}$ &  35.4~meV \\
  $\Delta_{\rm SO}$ & 0.341~eV & $E_g$ &  1.55~eV \\
  $\Delta E_{LT}$ & $0.08$~meV & $a_0$ & $146.1$~\AA \\
  $g_e$ & $-0.44$ & $m^{\ast}$ & $0.067m_0$ \\
  $g_{\rm Mn}$ & 2 & $S$ & 5/2 \\
  \hline \hline
\end{tabular}
\end{center}
\label{parameter}
\end{table}

\subsection{Spin relaxation in $n$-type Ga(Mn)As quantum wells}

In this subsection, we study the electron spin relaxation in $n$-type
Ga(Mn)As quantum wells, where the Mn dopants mainly take interstitial
positions. For simplicity, we neglect the substitutional Mn's and
assume that electrons from Mn donors are all free electrons.
We will discuss the situations that the quantum wells are either undoped or
$n$-doped before Mn-doping.

\begin{figure}[htbp]
  \begin{center}
    \includegraphics[height=6cm]{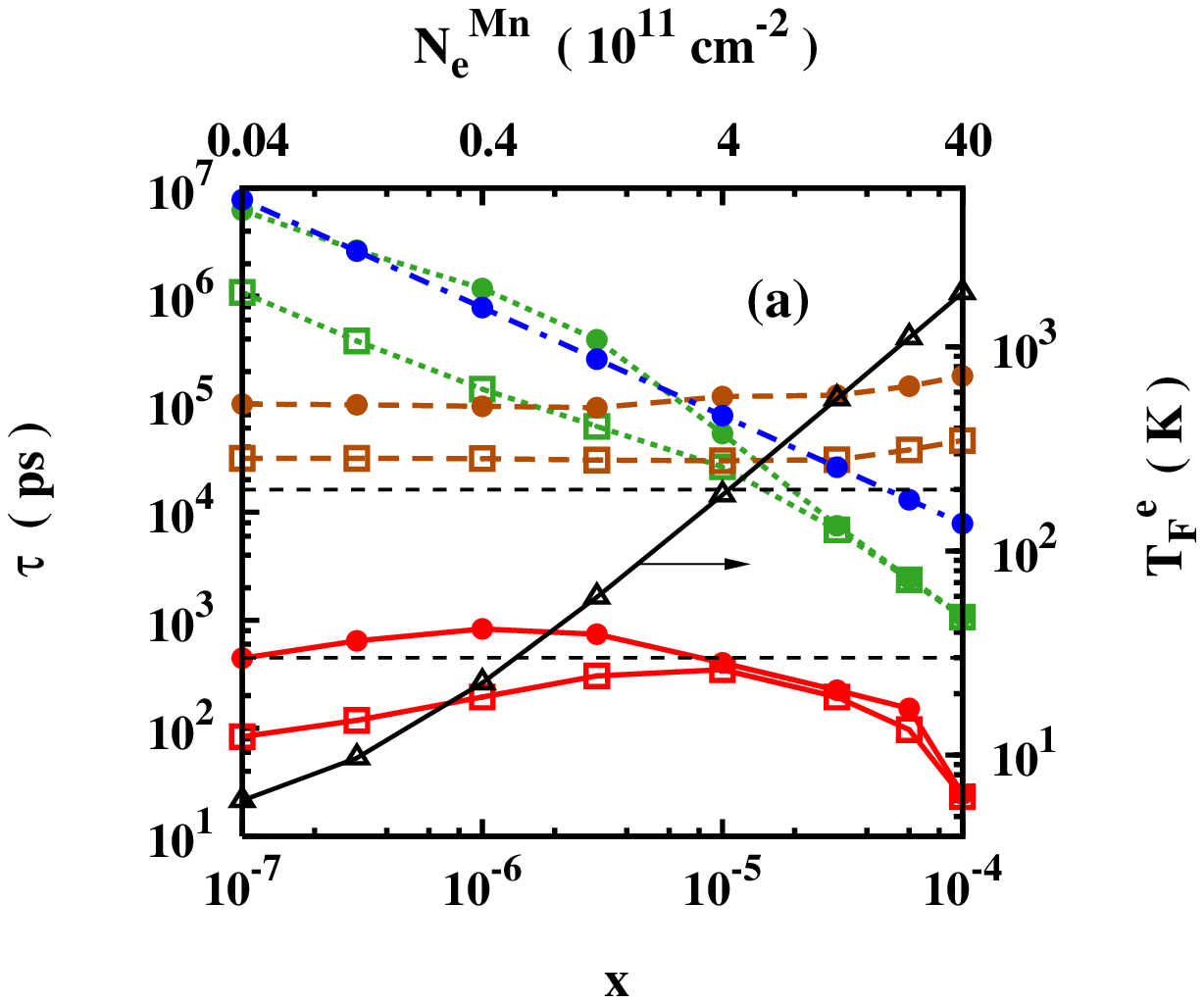}
  \end{center}
  \begin{center}
    \includegraphics[height=6cm]{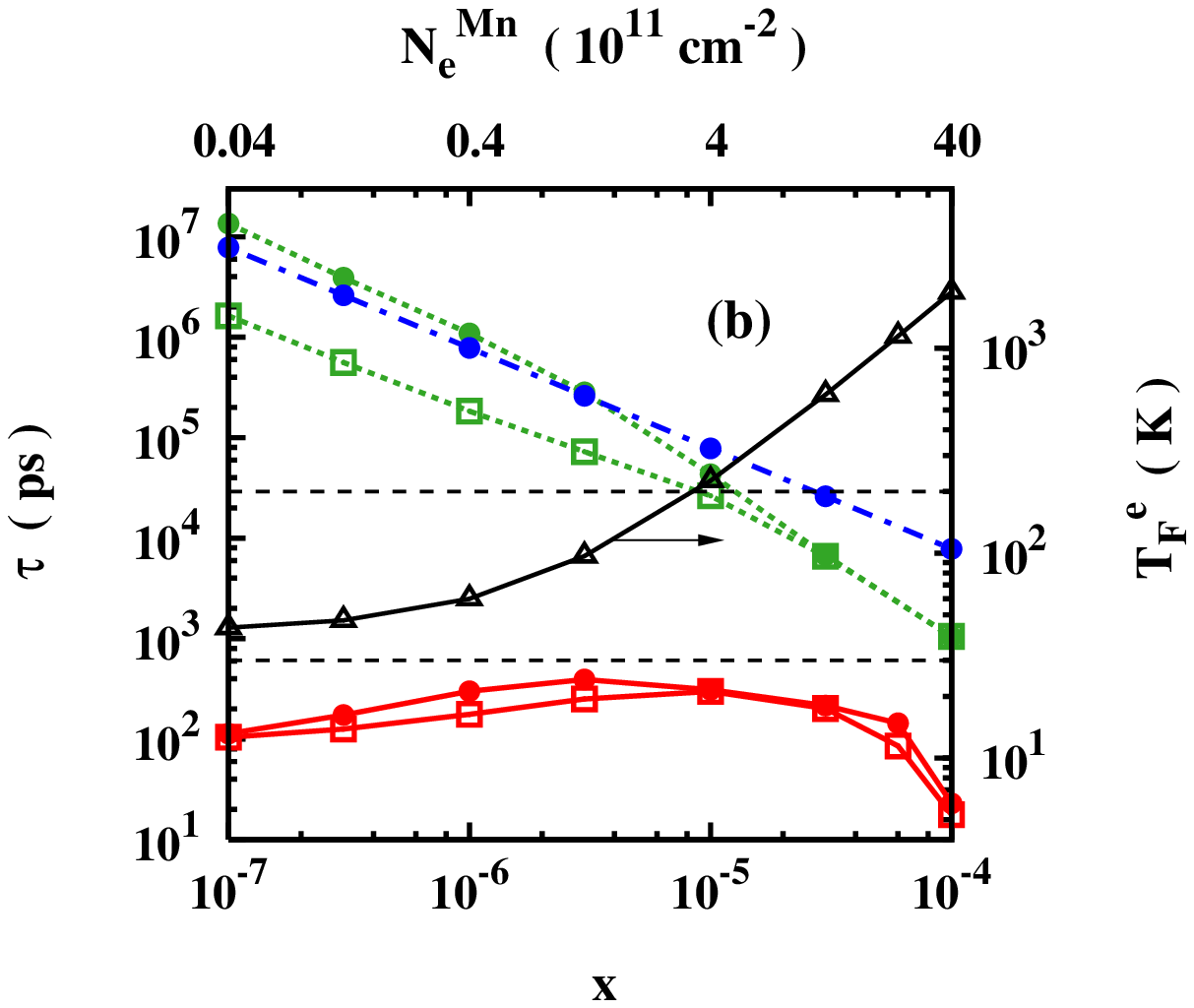}
  \end{center}
  \caption{ (Color online) SRT $\tau$ due to various mechanisms in
$n$-type Ga(Mn)As quantum wells which are (a) undoped or (b) $n$-doped
before Mn-doping as function of Mn concentration $x$ at 30~K ($\bullet$)
    and 200~K ($\square$). Red solid curves: the SRT due to the DP
    mechanism $\tau_{\rm DP}$; Green dotted curves: the SRT due to the
EY mechanism $\tau_{\rm EY}$; Brown dashed curves: the SRT due to
    the BAP mechanism $\tau_{\rm BAP}$; Blue chain curve: the SRT due
    to the $s$-$d$ exchange scattering mechanism $\tau_{\rm sd}$.
    The Fermi temperature of electrons $T_{\rm F}^e$ is ploted as black
    curve with $\triangle$ (the scale of $T_{\rm F}^e$ is on the right hand
    side of the frame) and $T_{\rm F}^e=T$ for both $T=30$ and $200$~K
    cases are plotted as black dashed curves. We also plot the scale
    of the electron density from Mn donors $N_e^{\rm Mn}$ on the
 top of the frame.}
  \label{n-itype}
\end{figure}

For quantum wells which are undoped before Mn-doping,
$N_e=N_e^{\rm Mn}+N_{\rm ex}$ where $N_e^{\rm Mn}$ is the density of
electrons from Mn donors and $N_{\rm ex}$ is the photo-excitation
density. We choose $N_{\rm ex}=10^{10}$~cm$^{-2}$ which is usually
smaller than $N_e^{\rm Mn}$. The SRTs due to various mechanisms are
plotted as function of $x$ in Fig.~\ref{n-itype}(a). $N_0\alpha$ is
chosen to be $-0.25$~eV, which is smaller (i.e., the $s$-$d$ exchange
interaction is stronger) than the value measured by
experiments.\cite{Awsch} However, even for such a strong exchange coupling,
the spin relaxation due to the $s$-$d$ exchange scattering mechanism
is {\it still} much weaker than that due to the DP mechanism. It
is further seen from
Fig.~\ref{n-itype}(a) that the BAP and EY mechanisms are also
unimportant. Therefore, the SRT is determined by the DP
mechanism. Interestingly, the SRT due to the DP mechanism
$\tau_{\rm DP}$ first increases then decreases with increasing $x$. The
$\tau$-$x$ curve thus has a peak. The underlying physics is that the
SRT has different $x$ (density) dependence in the non-degenerate and
degenerate regimes. Similar behavior has been found in bulk
non-magnetic III-V semiconductors in
Ref.~\onlinecite{bulk} very recently. Let us first recall the widely
used expression, $\tau_{\rm DP}=1/[\langle |{\bf h}({\bf k})|^2\rangle
  \tau_p]$ ($\langle...\rangle$ denotes the ensemble average), which is
derived within the elastic scattering approximation and is only
correct qualitatively.\cite{bulk} The expression contains two key
factors of the DP spin relaxation: (i) the inhomogeneous broadening
from the ${\bf k}$-dependent spin-orbit field $\sim \langle |{\bf
h}({\bf k})|^2\rangle$; (ii) the momentum scattering time $\tau_p$.
The SRT due to the DP mechanism increases with increasing momentum scattering
but decreases with increasing inhomogeneous broadening. It should be mentioned
that for this system, $N_e\approx n_i/2 = 2 x N_0$ (Note that
the charge number $Z$ of the Mn ion is included in $n_i$ as $Z^2$.
For interstitial Mn, which acts as a double donor, $Z=2$.) In the
small $x$ (low density) regime, the electron system is in the
non-degenerate regime, and the distribution is close to the
Boltzmann distribution. Therefore, the inhomogeneous broadening of
the ${\bf k}$ dependent spin-orbit field $\sim \langle |{\bf h}({\bf
k})|^2\rangle$ changes little with electron density $N_e$ (hence
$x$). On the other hand, in the non-degenerate regime the
electron-electron scattering increases with increasing electron
density\cite{Harley,Vignale} (thus $x$). Moreover, the
electron-impurity scattering also increases with increasing $x$ as the impurity
density increases. Therefore, $\tau_{\rm DP}$ increases with increasing $x$
(motional narrowing). In large $x$ (high density) regime, the
electron system is in the degenerate regime, the inhomogeneous
broadening changes as $k_{\rm F}^2\propto N_e\propto x$ ($k_{\rm F}^6\propto
N_e^3\propto x^3$) if the linear (cubic) term dominates the SOC. On
the other hand, the electron-electron scattering decreases with increasing
electron density (thus $x$) in the degenerate
regime.\cite{Harley,Vignale} Besides, the electron-impurity
scattering increases slower than $N_i\propto x$ because the
scattering cross section decreases as the electron (Fermi) energy
increases. Thus for both the linear- and cubic-term dominant cases,
$\tau_{\rm DP}$ decreases with increasing $x$ in the large $x$ regime. Consequently,
$\tau_{\rm DP}$ first increases then decreases with increasing $x$ and a peak
is formed in the crossover regime where $T\sim T_{\rm F}^e$ ($T_{\rm F}^e$ is
the electron Fermi temperature). It is seen from
Fig.~\ref{n-itype}(a) that
 for both $T=30$ and $200$~K cases, the peaks indeed appear at
$T\sim T_{\rm F}^e$. It should be pointed out that the situation here is
 different from that in Ref.~\onlinecite{Lai}, where the density
 dependence of the SRT also has a peak in intrinsic quantum wells at room
temperature. In that case, the impurity density is rather low and
the relevant momentum scatterings are the carrier-carrier
Coulomb and electron-phonon scatterings. In the situation here, the
impurity density is extremely high ($n_i=2N_e$) and the relevant
momentum scattering is the electron-impurity scattering.

We now turn to the situation that the quantum wells are $n$-doped
before Mn-doping. In this case, $N_e=N_e^{i}+N_e^{\rm Mn}+N_{\rm ex}$, where
$N_e^{i}$ denotes the density of the electrons from other dopants
which is chosen to be $10^{11}$~cm$^{-2}$. We assume that the other
dopants are far away from the quantum wells, so that they contribute
little to the electron-impurity scattering, corresponding to the
genuine case of modulation doping. However, the Mn ions are doped in
the quantum wells.\cite{Schulz,Korn} The photo-excitation density
  is $N_{\rm ex}=10^{10}$~cm$^{-2}$. The results are plotted in
Fig.~\ref{n-itype}(b). As the density of the photo-excited holes is much
smaller than the electron density, the BAP mechanism is obviously
negligible and thus not plotted in the figure. From the figure,
it is noted that the EY and $s$-$d$ exchange scattering mechanisms are
also insignificant. Consequently, the spin relaxation is still
dominated by the DP mechanism. Similar to that in Fig.~\ref{n-itype}(a),
the SRT due to the DP mechanism
$\tau_{\rm DP}$ first increases then decreases with increasing
$x$. For the case of $T=200$~K, the peak of the SRT is still around
$T=T_{\rm F}^e$. However, for the case of $T=30$~K, the peak moves to a larger
$x$ value compared to that in Fig.~\ref{n-itype}(a). This can be
understood by noting that the electrons have two sources: the Mn
donors and other dopants. For the case of $T=200$~K, the
crossover of the non-degenerate and degenerate regimes takes place
around $T\sim T_{\rm F}^e$, where the corresponding $x$ is $10^{-5}$. At
such $x$, electrons are mainly from the Mn ions rather than from
other dopants. Thus the situation is the same as that in
Fig.~\ref{n-itype}(a) and the peak appears at $T\sim T_{\rm F}^e$. However, in
the case of $T=30$~K, for all $x$ in the figure, $T_{\rm F}^e$ is larger than
$T$ and the situation is hence different. The $\tau$-$x$ behavior in
this case can be understood as follows: For $x< 10^{-6}$,
electrons are mainly from the other dopants and $N_e$ changes slowly
with $x$, thus the inhomogeneous broadening varies slowly with
$x$. On the other hand, the electron-impurity scattering increases
as the impurity density increases $n_i \approx 4xN_0$. At low
temperature ($T<T_{\rm F}^e$), the
electron-impurity scattering can be important even when
$n_i<N_e$.\cite{lowT,bulk} Therefore, the momentum scattering increases
with $x$ significantly. Consequently, $\tau_{\rm DP}$ increases with increasing
$x$. For $x > 10^{-5}$, electrons mainly come from the Mn donors. The
scenario becomes the same as that in the case of Fig.~\ref{n-itype}(a)
and the SRT decreases with increasing $x$ as $T_{\rm F}^e$ is much larger than
$T$. Consequently, the peak is formed in the range $10^{-6} < x
<10^{-5}$ at $x = 3 \times 10^{-6}$, which is larger than that in the
case of $T=30$~K in Fig.~\ref{n-itype}(a).

It should be mentioned that our results are consistent with the latest
experimental finding that in the low Mn concentration regime the SRT
increases with increasing $x$.\cite{Schulz,Korn}

\subsection{Electron spin relaxation in $p$-type Ga(Mn)As quantum wells}

\begin{figure}[htb]
  \begin{center}
    \includegraphics[height=6cm]{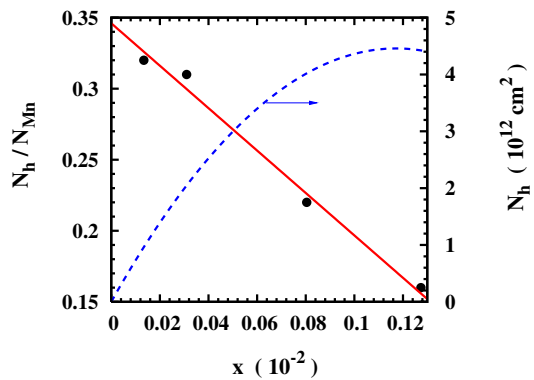}
  \end{center}
  \caption{ (Color online) Ratio of the hole density to the Mn
    density $N_h/N_{\rm Mn}$ {\em vs.} the Mn concentration $x$ in
    $p$-type Ga(Mn)As quantum wells. The black dots represent the
    experimental data. The red solid curve is the fitted one. The hole
    density $N_{h}$ is also plotted (the blue dashed curve). Note that
    the scale of $N_h$ is on the right hand side of the frame.}
  \label{n_h-ptype}
\end{figure}

In this subsection, we discuss the electron spin relaxation in
$p$-type Ga(Mn)As quantum wells. Both
substitutional and interstitial Mn ions exist in the system.
Each substitutional Mn donates one hole, whereas each interstitial
Mn compensates two holes. For simplicity, we assume that all the
holes are free. The ratio of the hole density $N_h$ to
the Mn density $N_{\rm Mn}$ is obtained by fitting the experimental
data in Ref.~\onlinecite{Awsch}, as shown in Fig.~\ref{n_h-ptype}.
From these densities, according to charge neutrality, the densities
of substitutional Mn $N_{\rm Mn}^{\rm S}$ and interstitial Mn
$N_{\rm Mn}^{\rm I}$ are determined. The photo-excitation density is
chosen to be $N_{\rm ex}=5\times 10^{10}$~cm$^{-2}$ unless otherwise
specified.

\begin{figure}[htb]
  \begin{center}
    \includegraphics[height=5.8cm]{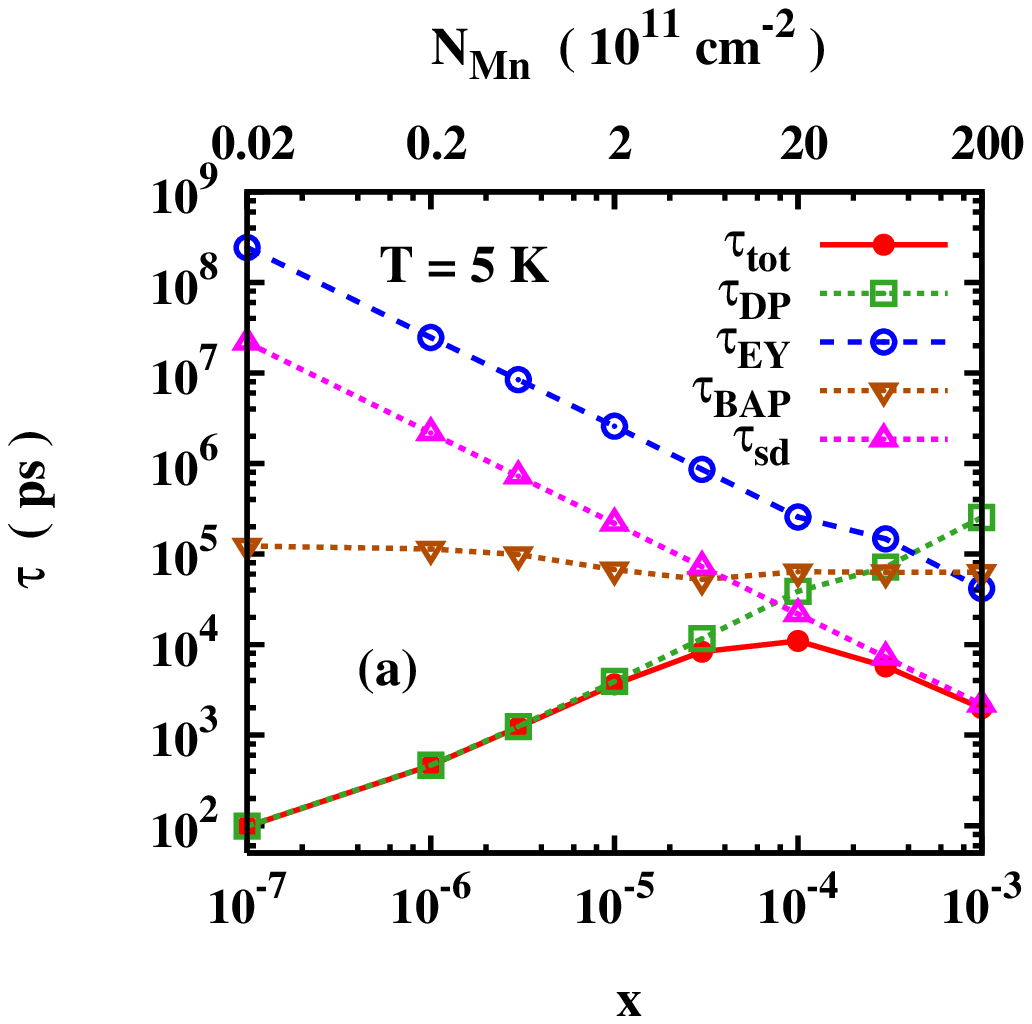}
  \end{center}
  \begin{center}
    \includegraphics[height=5.8cm]{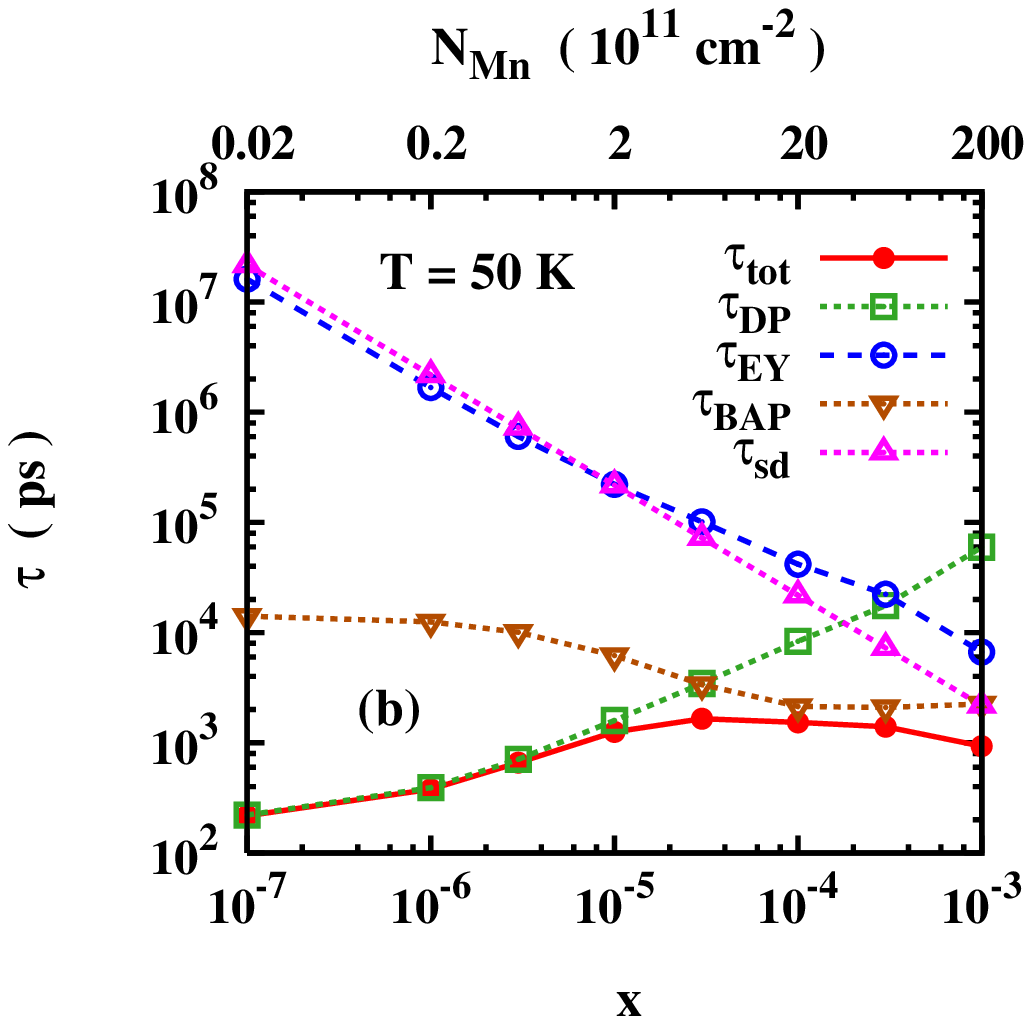}
  \end{center}
  \begin{center}
    \includegraphics[height=5.8cm]{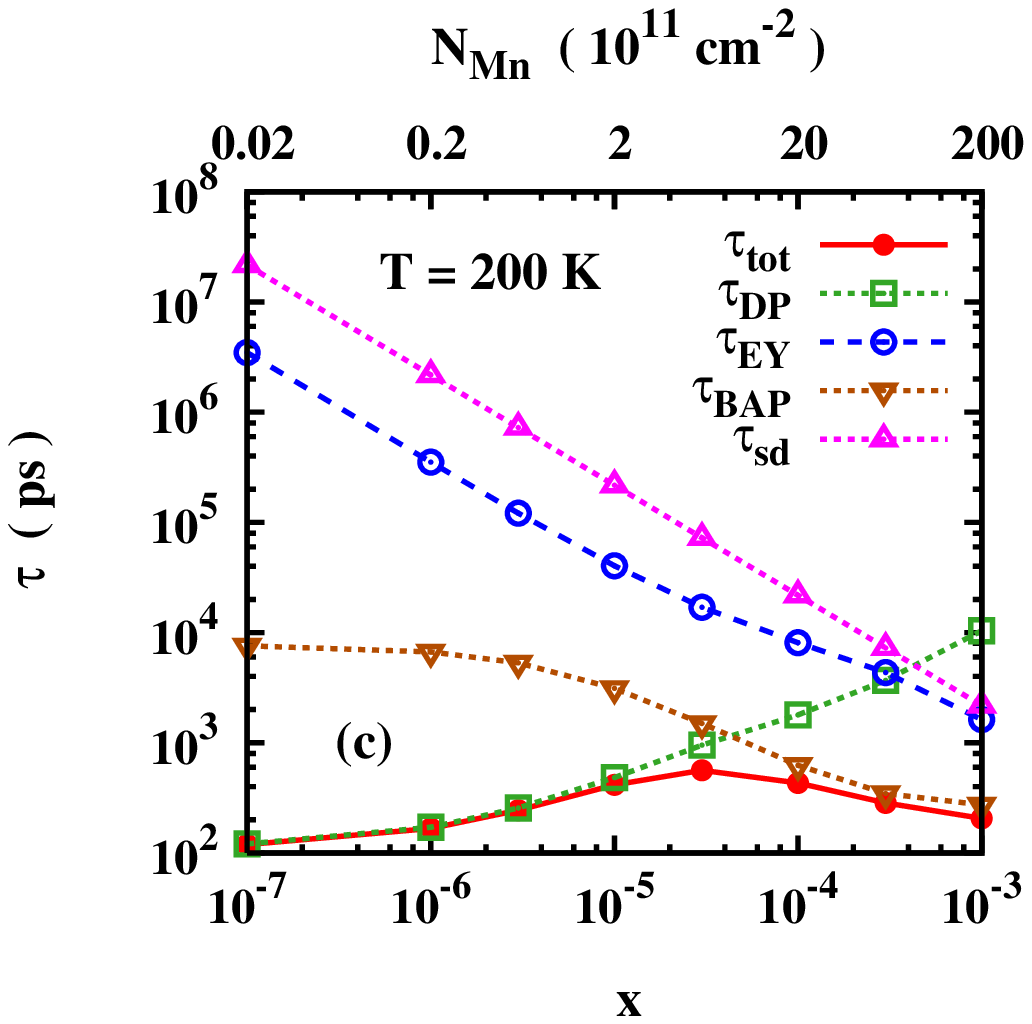}
  \end{center}
  \caption{ (Color online) SRT $\tau$ due to various mechanisms and
    the total SRT in $p$-type Ga(Mn)As against the Mn
    concentration $x$ at (a) $T=5$, (b) 50 and (c) $200$~K. We also plot
    the scale of $N_{\rm Mn}$ on the top of the frame.}
  \label{srt-x}
\end{figure}

\subsubsection{Mn concentration dependence of the SRT}

We first study the Mn concentration dependence of the SRT. In
Fig.~\ref{srt-x}, we plot the SRTs due to various mechanisms and the
total SRT as function of the Mn concentration $x$ at $T=5$, 50 and
$200$~K. It is noted that the total SRT first increases and then
decreases with increasing $x$ and there is a peak at $x\sim 3\times 10^{-5}$.
Remarkably, the spin relaxation at large $x$ is not dominated by the
DP mechanism, but by the $s$-$d$ exchange scattering (or the BAP)
mechanism at low (or high) temperature. At medium temperature, the
EY mechanism also contributes for large $x$. The SRT due to the DP
mechanism increases with increasing $x$, whereas those due to the $s$-$d$
exchange, EY and  BAP mechanisms decrease with increasing $x$. Consequently a peak is
formed. It should be pointed out that the underlying physics here is
different from that in the case of $n$-type Ga(Mn)As quantum well 
where the peak is solely due to the electron density dependence of the
DP spin relaxation. It should be mentioned that the peak position is
$x\sim 10^{-4}$, which is consistent with that observed in
Ref.~\onlinecite{Awsch}. 

\begin{figure}[tbp]
  \begin{center}
    \includegraphics[height=6cm]{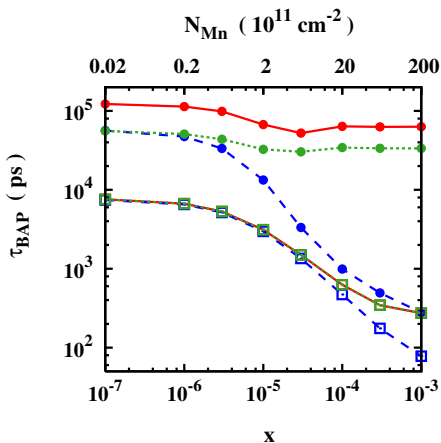}
  \end{center}
  \caption{ (Color online) SRT due to the BAP mechanism
    $\tau_{\rm BAP}$ as function of the Mn concentration at $T=5$
    ($\bullet$) and 200~K ($\square$). Red solid curves:
    $\tau_{\rm BAP}$ from the full calculation; Green dotted curves:
    $\tau_{\rm BAP}$ from the calculation without the Pauli blocking
    of electrons; Blue dashed curves: $\tau_{\rm BAP}$ from the calculation
    without the Pauli blocking of holes.}
  \label{srt-bap}
\end{figure}

Let us now turn to the $x$ dependence of the SRT due to various
mechanisms. The increase of $\tau_{\rm DP}$ with increasing $x$ is due to the
increase of the electron-impurity and electron-hole scatterings
(motional narrowing). For the EY mechanism, the SRT
decreases as the spin-flip scattering increases with increasing impurity density
[see Eq.~(\ref{equ:rho_EY_ei})]. The $s$-$d$ exchange scattering
increases with increasing $x$ as the Mn density increases [see
  Eq.~(\ref{equ:rho_ex_scat0})]. The $x$ dependence of the
the BAP spin relaxation is more complicated. To facilitate the
understanding, we plot $\tau_{\rm BAP}$ from the full calculation
and that from the calculations without the Pauli blocking of
electrons (holes) in Fig.~\ref{srt-bap}. It is seen that for
$x<10^{-6}$, $\tau_{\rm BAP}$ changes little with $x$.
This is due to the fact that the holes from
Mn dopants are much fewer than those from photo-excitation, and
hence $N_h$ changes little with $x$. So does $\tau_{\rm BAP}$. For
larger $x$, $\tau_{\rm BAP}$ first decreases then increases a little
and finally saturates with increasing $x$ at $T=5$~K. It is noted that without
the Pauli blocking of holes, $\tau_{\rm BAP}$ decreases with increasing $x$
rapidly, which indicates that the slowdown of the decrease and the
saturation of $\tau_{\rm BAP}$ are due to the Pauli blocking of
holes. It is further shown that the Pauli blocking of electrons is
also important as $T_{\rm F}^e=20$~K is larger than $T=5$~K. For the case
of $T=200$~K, the effect of the Pauli blocking of electrons is
negligible as $T\gg T_{\rm F}^e$. The Pauli blocking of holes becomes
visible only for $x>10^{-4}$, where the hole Fermi temperature
becomes larger than $T=200$~K.

From Eq.~(\ref{equ:rho_ex_scat0}), one can see that the spin
relaxation due to the $s$-$d$ exchange scattering is
independent of temperature. However, the spin relaxation due to the
BAP mechanism increases with increasing temperature because the Pauli blocking of
electrons and holes decreases with increasing temperature. Moreover, the matrix
element of the BAP mechanism increases with the center-of-mass
momentum of the interacting electron-hole pair, of which the ensemble average
hence increases with increasing temperature.\cite{wu-bap}
The spin relaxation due to the EY mechanism increases with increasing
temperature too, as the spin-flip matrices [$\hat{\Lambda}^{(1)}_{{\bf
      k}^\prime,{\bf k}}$ and $\hat{\Lambda}^{(2)}_{{\bf
      k}^\prime,{\bf k}}$] in Eq.~(\ref{equ:rho_EY_ei}) increase with increasing
$k$. Consequently, the BAP and EY mechanisms eventually become more
efficient than the $s$-$d$ exchange scattering mechanism at high
temperature.

The appearance of the peak in the $\tau$-$x$ curve has been observed
in a recent experiment at $5$~K.\cite{Awsch} However, the SRTs we
obtain are much larger than the experimental value
under the same conditions. The deviation may come from
pretermission of localized holes. At such low temperature ($5$~K),
the localization of holes is not negligible.\cite{localize,opt-or} The
localized holes act as exchange interaction centers located randomly
in the sample, which thus lead to spin relaxation similar to the
$s$-$d$ exchange scattering mechanism. As there is no Pauli blocking
of the localized holes, the spin relaxation can be very efficient.\cite{opt-or}
It should be mentioned that, however, recent studies have also shown
that there is some compensation of the $s$-$d$ exchange interaction
and the electron-hole exchange interaction as holes are always
localized on the Mn acceptors.\cite{Astakhov,Sliwa} This leads to a
longer spin relaxation time\cite{Astakhov} and smaller measured (by
magneto-optical techniques) $s$-$d$ exchange coupling constant\cite{Sliwa}. 
However, for the high temperature case, the localization is marginal
and our consideration is close to the genuine case. The predicted
$\tau$-$x$ dependence should be tested experimentally.

\subsubsection{Temperature dependence of the SRT}

\begin{figure}[htb]
  \begin{center}
    \includegraphics[height=6cm]{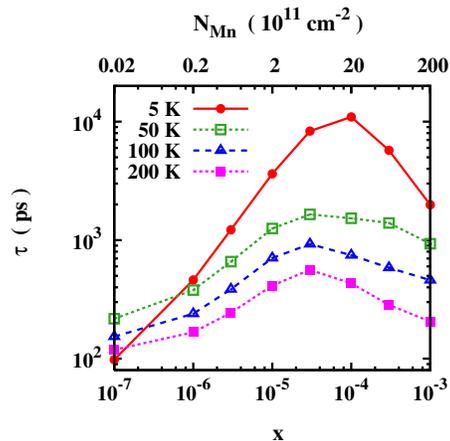}
  \end{center}
  \caption{ (Color online) SRT $\tau$ as function of Mn concentration
    $x$ at different temperatures.}
  \label{srt-T1}
\end{figure}

\begin{figure}[htb]
  \begin{center}
    \includegraphics[height=6cm]{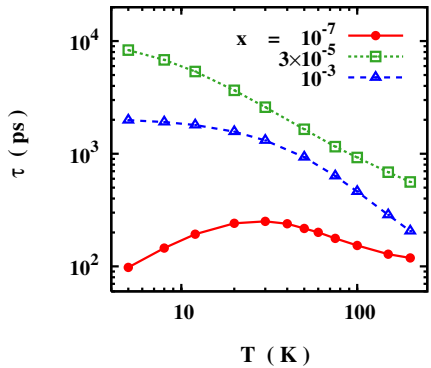}
  \end{center}
  \caption{ (Color online) SRT $\tau$ as function of temperature $T$
    for different Mn concentrations.}
  \label{srt-T2}
\end{figure}

We now discuss the temperature dependence of the SRT. In
Fig.~\ref{srt-T1}, we plot the SRT as function of Mn concentration $x$
for different temperatures. It is seen that for each case the
$\tau$-$x$ curve shows a peak. It is further noted that the temperature
dependences of the SRT are different for small (e.g., $x=10^{-7}$)
and large $x$ (e.g., $x=10^{-3}$). To make it more pronounced,
we further plot the temperature
dependence of the SRT for $x=10^{-7}$, $3\times 10^{-5}$ and $10^{-3}$ in
Fig.~\ref{srt-T2}. For $x=10^{-7}$, the SRT first increases then
decreases with increasing temperature and there is a peak around $20$~K. It is
understood that for such a small $x$, the electrons and holes are
mainly from the photo-excitation. For such system, the
electron-electron and electron-hole Coulomb scatterings are most
important. It is shown in Ref.~\onlinecite{lowT} that the nonmonotonic
temperature dependence of the electron-electron Coulomb scattering
leads to a peak in the $\tau$-$T$ curve. In the situation here, the
electron-hole Coulomb scattering also contributes to the formation of
the peak. For the case of $x=3\times 10^{-5}$, all spin relaxation mechanisms are
  relevant and the most important momentum scattering is
the electron-impurity scattering. In this case the SRT due to the DP
mechanism decreases with increasing temperature monotonically as the
increase of the inhomogeneous broadening dominates.\cite{lowT}
Moreover, the SRTs due to the EY and BAP mechanisms also decrease with
increasing temperature. Consequently, the total SRT decreases
with increasing temperature monotonically. For the case of large
$x$ ($x=10^{-3}$), the spin relaxation is dominated by the $s$-$d$
exchange scattering (or the BAP) mechanism at low (or high)
temperature. As the $s$-$d$ exchange scattering mechanism is
independent of the temperature, the temperature dependence is rather
weak in the low temperature regime. As the temperature increases, the
EY and BAP mechanisms become more and more important, which leads to a
fast decrease of the SRT with temperature.

\subsubsection{Photo-excitation density dependence of the SRT}

\begin{figure}[htb]
  \begin{center}
    \includegraphics[height=6cm]{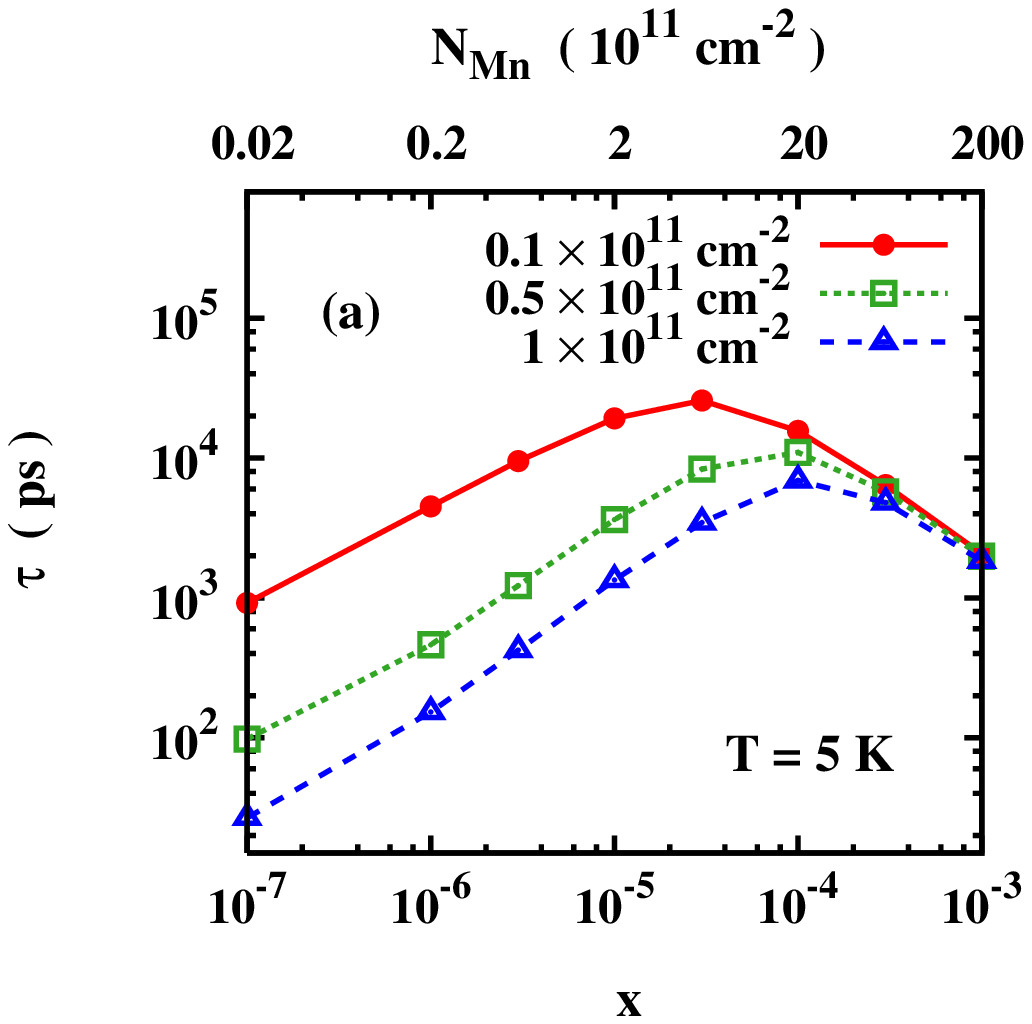}
  \end{center}
  \begin{center}
    \includegraphics[height=6cm]{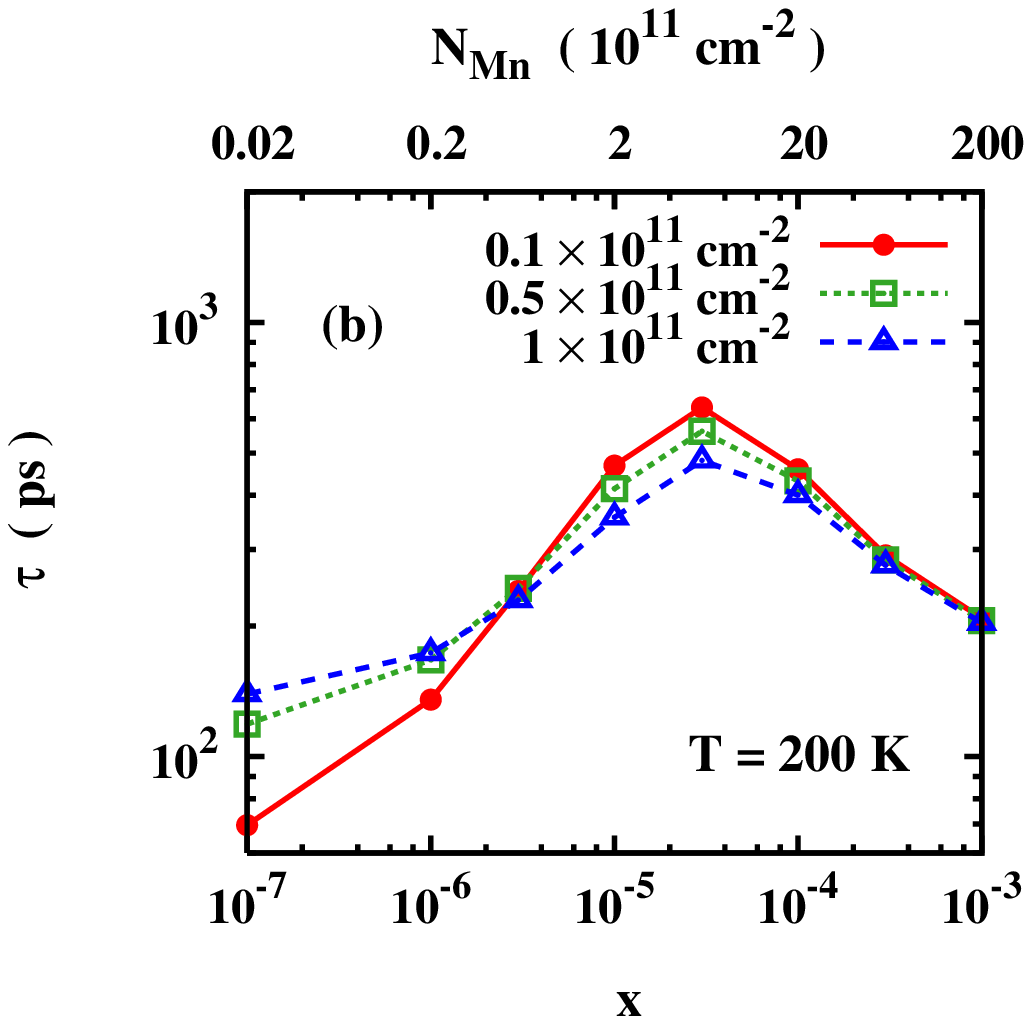}
  \end{center}
  \caption{ (Color online) SRT $\tau$ as function of the Mn
    concentration for different photo-excitation densities at (a)
    $T=5$ and (b) $200$~K. Red solid curve with $\bullet$:
    $N_{\rm ex}=0.1\times 10^{11}$~cm$^{-2}$; Green dotted curve with $\square$:
    $N_{\rm ex}=0.5\times 10^{11}$~cm$^{-2}$; Blue dashed curve with
    $\triangle$: $N_{\rm ex}=1\times 10^{11}$~cm$^{-2}$.}
  \label{srt-Nex}
\end{figure}

We now study the photo-excitation density $N_{\rm ex}$ dependence of the
spin relaxation. In Fig.~\ref{srt-Nex} the SRT is plotted against
the Mn concentration $x$ for three photo-excitation densities at low
(5~K) and high (200~K) temperatures. It is noted that the SRT
exhibits very different photo-excitation density dependences at low
and high temperatures. Moreover, the photo-excitation density
dependence varies with $x$. Let us divide the variation of $x$ into
three regimes: the small $x$ ($x<3\times 10^{-6}$) regime where the
DP mechanism is dominant; the medium $x$ ($3\times
10^{-6}<x<10^{-4}$) regime where the DP mechanism is comparable with
the other mechanisms; the large $x$ ($x>10^{-4}$) regime where the
DP mechanism is irrelevant. 

In small $x$ regime, the DP mechanism
dominates the spin relaxation. The photo-excitation dependence of
the DP spin relaxation is different in the degenerate and
non-degenerate regimes. Similar to the case of $n$-type Ga(Mn)As
quantum wells (see Sec.~IIIA), in degenerate (low temperature)
regime, the density dependence of the SRT is dominated by the
increase of the inhomogeneous broadening with increasing density, and hence the
SRT decreases with increasing density. In non-degenerate (high temperature)
regime, the density dependence of the SRT is dominated by the
increase of the electron-electron and electron-hole Coulomb
scatterings with density, and hence the SRT increases. 

In the large $x$ regime, the spin relaxation is mainly due to the EY, BAP and
$s$-$d$ exchange scattering mechanisms. At low temperature, the
$s$-$d$ exchange scattering mechanism is dominant. As $\tau_{\rm sd}$ is
independent of the electron density, the photo-excitation density
dependence of the SRT (which mainly comes from the EY
mechanism) is weak. At high temperature, the BAP mechanism
dominates. As holes are mainly from the Mn dopants and the electron
system is non-degenerate, the SRT also changes little with
photo-excitation density. 

In the medium $x$ regime, all the four
mechanisms contribute to the spin relaxation. As $\tau_{\rm sd}$ is
independent of the electron density, the density dependence comes
from the other three mechanisms. At low temperature, besides
the DP mechanism, the BAP mechanism is also important. However, the
BAP spin relaxation changes slowly with electron (hole) density as
the Pauli blocking is important at low temperature (see
Fig.~\ref{srt-bap}). The spin relaxation due to the EY mechanism
also increases with increasing $N_{\rm ex}$ as the spin-flip matrices
[$\hat{\Lambda}^{(1)}_{{\bf
      k}^\prime,{\bf k}}$ and $\hat{\Lambda}^{(2)}_{{\bf
      k}^\prime,{\bf k}}$] in Eq.~(\ref{equ:rho_EY_ei}) increase with increasing
  $k$. However, the EY mechanism is usually less efficient than the DP
mechanism for $x< 10^{-4}$.
As the DP spin relaxation increases with increasing photo-excitation density
while other relevant mechanisms change slowly or less important
  than it, the peak moves to
larger $x$ with increasing photo-excitation density as indicated in
Fig.~\ref{srt-Nex}(a). Moreover, the total SRT decreases with increasing
photo-excitation density.
At high temperature, as the electron system is non-degenerate,
the inhomogeneous broadening changes slowly with
  $N_{\rm ex}$. However, the screening increases with increasing $N_{\rm ex}$ as the
  carrier density increases. Hence the
  momentum scattering (mainly from the electron-impurity scattering)
  decreases with increasing $N_{\rm ex}$. Therefore, the SRT due to the DP
 mechanism decreases with increasing $N_{\rm ex}$. The EY mechanism is
less important than the DP mechanism in this regime [see
    Fig.~\ref{srt-Nex}(c)]. Moreover, as the hole system is
non-degenerate, the spin relaxation due to the BAP mechanism increases
with hole density. Therefore, the SRT also decreases with increasing
photo-excitation density.

\subsubsection{Effect of magnetic field on the SRT}

We now study the effect of magnetic field on the SRT. The magnetic
field is applied parallel to the quantum well plane, which is
perpendicular to the initial electron spin polarization (the Voigt
configuration). In traditional non-magnetic $n$-type quantum wells,
where the electron spin relaxation is dominated by the DP mechanism,
the magnetic field in the Voigt configuration has dual effects on spin
relaxation: (i) elongating the spin lifetime by a factor of
$[1+(\omega_L \tau_p)^2]$ ($\omega_L$ is the Larmor frequency,
$\tau_p$ is the momentum scattering time);\cite{opt-or} (ii) mixing
the in-plane and out-of-plane spin relaxations,\cite{Dohrmann,highP} e.g.,
$\frac{1}{\tau}=\frac{1}{2}(\frac{1}{\tau_{z}} + \frac{1}{\tau_{\|}})$
when $\omega_L>\frac{1}{2}(\frac{1}{\tau_{z}} -
\frac{1}{\tau_{\|}})$.\cite{Dohrmann} Usually, effect (i) is weak,
but effect (ii) is more important. Differing from the case of
non-magnetic $n$-type quantum wells, there are several new scenarios
in the $p$-type Ga(Mn)As quantum wells: (i) The magnetic field can
polarize the Mn spins, which alters the spin relaxation due to
the $s$-$d$ exchange scattering mechanism. (ii) The nonequilibrium Mn
spin polarization can be induced during the evolution through the
$s(p)$-$d$ exchange interaction with both electrons and holes. It precesses
around the magnetic field and produces the Mn beats. This has been
studied both experimentally and theoretically in II-VI magnetically
doped quantum wells.\cite{Linder,Crooker}
However, we find that the induced nonequilibrium Mn spin polarization
is rather small (much smaller than the electron spin polarization) and
affects the spin dynamics marginally. This is
consistent with the fact that the Mn beats are not observed in $p$-type
Ga(Mn)As quantum wells.\cite{Awsch} (iii) the spin relaxation
due to the EY mechanism is also anisotropic because the out-of-plane spin
relaxation comes from $\hat{\Lambda}^{(1)}_{{\bf k}^\prime,{\bf k}}$
while the in-plane relaxation from $\hat{\Lambda}^{(2)}_{{\bf k}^\prime,{\bf
    k}}$ [see Eq.~(\ref{equ:rho_EY_ei})]. At large $x$ when the EY
mechanism is important, this anisotropy may show up.

\begin{figure}[htb]
  \begin{center}
    \includegraphics[height=5.2cm]{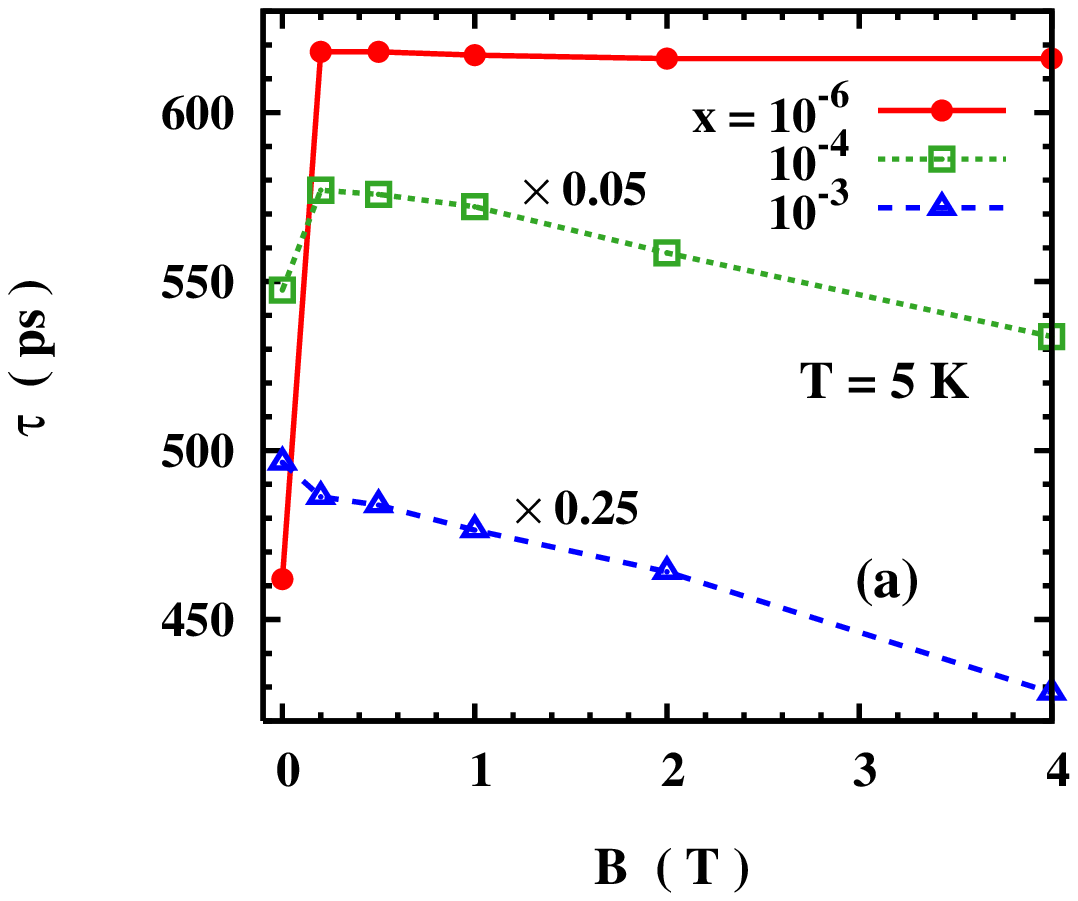}
  \end{center}
  \begin{center}
    \includegraphics[height=5.2cm]{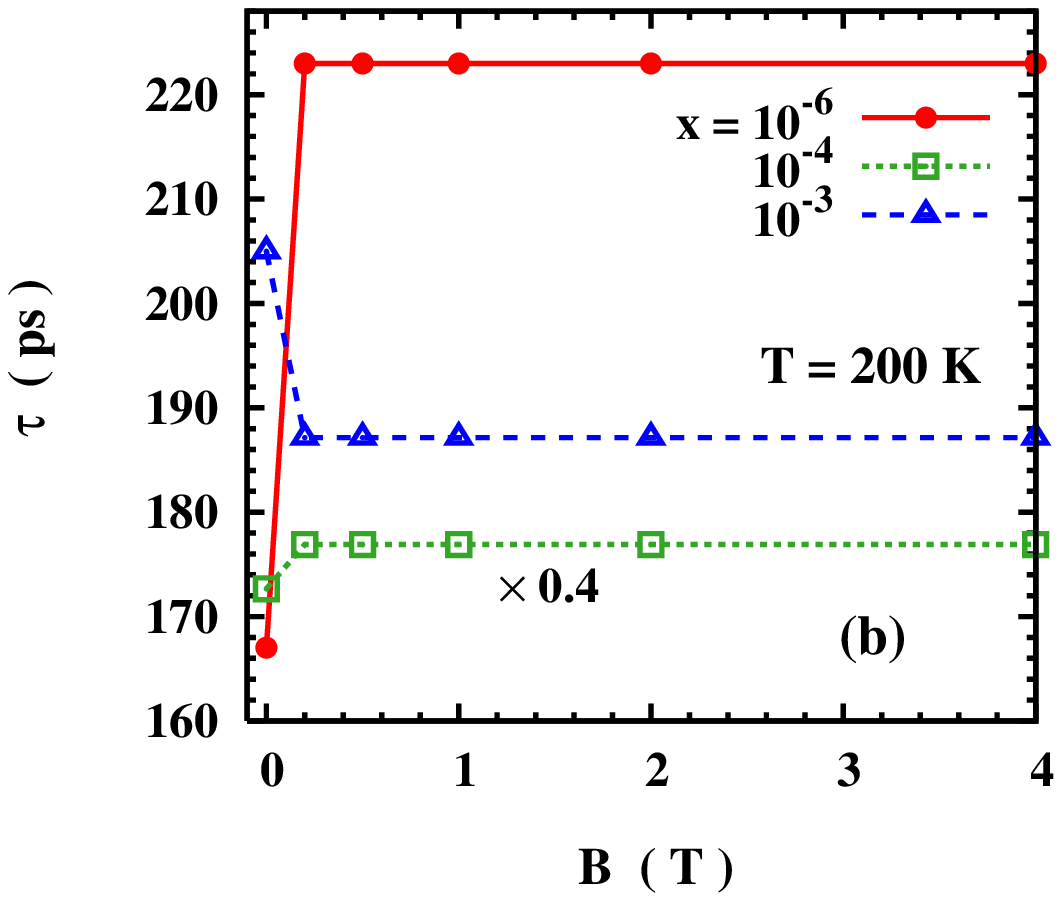}
  \end{center}
  \caption{(Color online) SRT $\tau$ {\em vs}. the magnetic
    field with different Mn concentrations at (a) $T=5$~K and (b)
    $200$~K. Note that the values in the figure have been rescaled by
    a factor of 0.25 for the case of $x=10^{-3}$ at $T=5$~K, and 0.05
    (0.4) for the case of $x=10^{-4}$ at $T=5$ (200)~K.}
  \label{srt-B}
\end{figure}

In Fig.~\ref{srt-B}, we plot the SRT as function of the
magnetic field with different Mn concentration at $T=5$ and 200~K.
For the case of small $x$ ($x=10^{-6}$), it is seen that the SRT
increases abruptly when the magnetic field varies from 0 to 0.2~T and
is almost a constant for $B=0.2$ to 6~T. This abrupt increase of the SRT
originates from the mixing of the out-of-plane and in-plane electron spin
relaxations in the presence of magnetic field. For small $x$, the spin
relaxation is dominated by the DP mechanism. For the DP spin
relaxation, the in-plane spin relaxation is slower than the out-of-plane
one, as only part of the inhomogeneous spin-orbit field ${\bf h}({\bf
  k})$ contributes to the in-plane spin relaxation. After the magnetic
field is applied, the spin relaxation rate becomes
$\frac{1}{\tau}=\frac{1}{2}(\frac{1}{\tau_{z}} +
\frac{1}{\tau_{\|}})$. The condition for this relation is
$\omega_L>\frac{1}{2}(\frac{1}{\tau_{z}} - \frac{1}{\tau_{\|}})$. In
the situation considered here, it is $B\gtrsim 0.005$~T (0.02~T) for
low (high) temperature case. Therefore, the variation of the SRT with
the magnetic field seems abruptly.

For the case of large $x$ ($x=10^{-3}$), the relevant spin relaxation
mechanisms at high temperature are the BAP, EY and $s$-$d$ exchange
mechanisms. The BAP and $s$-$d$ exchange mechanisms are isotropic.
However, the EY mechanism is anisotropic.
 Our calculation indicates that the 
$\hat{\Lambda}^{(1)}_{{\bf k}^\prime,{\bf k}}$ term
is smaller than the $\hat{\Lambda}^{(2)}_{{\bf k}^\prime,{\bf k}}$
term in Eq.~(\ref{equ:rho_EY_ei}). Hence the in-plane EY spin relaxation
is faster than the out-of-plane one.
%\Green{
% and the SRT decreases with magnetic field abruptly}.
Therefore the SRT decreases
  with increasing magnetic field abruptly by a small amount at low magnetic field. 
After the abrupt decrease, the SRT changes little with the
magnetic field as the BAP mechanism is almost independent of the
magnetic field. At low temperature, the dominant spin
relaxation mechanism is the $s$-$d$ exchange scattering mechanism 
which is isotropic. Moreover, the contribution of the EY
  mechanism is even smaller compared to the high temperature case [see
    Fig.~\ref{srt-x}(a) and (c)]. Therefore, the magnitude of the abrupt
decrease of the SRT at low magnetic field is even smaller than the
high temperature case.
Let us now turn to the magnetic field dependence of the
$s$-$d$ exchange scattering mechanism. We choose the eigenstates of
$\sigma_x$ (denoted as $|\pm\rangle$) as
the basis [hence $\langle s_z \rangle=-\sum_{\bf k} \mbox{Im}\rho_{\bf
    k}$]. By keeping only the diagonal element of the Mn spin density
matrix $\hat{\rho}_{\rm Mn}$, from Eq.~(\ref{equ:rho_ex_scat0}) we obtain
\begin{eqnarray}
  \nonumber
  &&\mspace{-40mu} \partial_t \sum_{\bf k} \rho_{{\bf k}}\bigg|^{scat}_{\rm sd}
  =- N_{\rm Mn} \alpha^2 I_s \sum_{{\bf k}} \rho_{\bf k} \Big\{
  \frac{m^{\ast}}{4} \big[ S(S+1)
  \\ && \mbox{} \mspace{0mu}
    + \langle S_x^2 \rangle \big]
    - \sum_{{\bf k}^{\prime}}
    \pi \delta(\varepsilon_{{\bf k}^{\prime}}-\varepsilon_{{\bf k}} )
    \langle S_x \rangle (f_{{\bf k}^{\prime}+}-f_{{\bf k}^{\prime}-})
    \Big\}.
  \label{equ:rho_ex_scat3}
\end{eqnarray}
As $(f_{{\bf k}^{\prime}+}-f_{{\bf k}^{\prime}-})$ corresponds to the
electron spin polarization along the $x$-axis which is much smaller than
the Mn spin polarization as both the spin and the $g$-factor of the Mn
ions are larger than those of electrons. Therefore, the second
term in the right hand side of the above equation is much smaller than
the first one. The spin relaxation due to the $s$-$d$ exchange scattering
mechanism increases with increasing magnetic field as $\langle S_x^2
\rangle$ does. Consequently, after the abrupt decrease of the SRT
at low magnetic field, the SRT 
further decreases with increasing
  magnetic field due to the enhancement of the $s$-$d$ exchange
  scattering.

We now turn to the medium $x$ case ($x=10^{-4}$). At low
temperature, all the mechanisms are relevant
[see Fig.~\ref{srt-x}(a)]. As the BAP and $s$-$d$ exchange scattering
mechanisms are isotropic, the anisotropy mainly comes from the DP
and EY mechanisms. However, as the EY mechanism is less efficient
than the DP mechanism, the anisotropy mainly comes from the DP
one. Consequently, the SRT first increases abruptly due to
the mixing of the in-plane and out-of-plane DP spin relaxations, and then
decreases as the $s$-$d$ exchange scattering increases with increasing
magnetic field. At high temperature, the $s$-$d$ exchange scattering
mechanism is negligible. Hence after the abrupt increase, the SRT
varies little with the magnetic field.

\section{Conclusion}

In summary, we have performed a systematic investigation on the spin
relaxation in paramagnetic Ga(Mn)As quantum wells from a fully
microscopic KSBE approach with all the relevant scatterings explicitly
included.

For $n$-type Ga(Mn)As quantum wells, where most Mn ions take the
interstitial positions,\cite{Schulz,Korn,Edmonds} we find that the
spin relaxation is always dominated by the DP mechanism in the
metallic regime. Interestingly, the Mn concentration dependence of
the SRT is nonmonotonic and exhibits a peak. This
is due to the fact that the momentum scattering
and the inhomogeneous broadening have different density dependences
in the non-degenerate and degenerate regimes. A similar effect
was found in bulk III-V semiconductors very recently.\cite{bulk} Our
results also are consistent with the latest experimental
 finding that in the low Mn concentration regime the SRT increases
  with Mn concentration.\cite{Schulz,Korn}

For the $p$-type Ga(Mn)As quantum wells, we study the SRT for
various Mn concentrations, temperatures, photo-excitation densities
and magnetic fields. It is found that the SRT first increases then
decreases with increasing Mn concentration. The underlying physics is as
follows: In the regime of small Mn concentration $x$, the spin
relaxation is dominated by the DP mechanism which decreases with
increasing impurity (Mn) density (hence $x$) due to motional
narrowing; In the large $x$ regime, as the Mn and hole
densities are very large, the spin-flip scatterings such as the EY
mechanism associated with the electron-impurity scattering, the
$s$-$d$ exchange scattering and the electron-hole exchange
scattering become more important than the DP spin relaxation. The
SRT hence decreases with increasing Mn concentration $x$ and the peak is
formed. It is found that the most important spin relaxation
mechanism at large $x$ is the $s$-$d$ exchange scattering (or the
BAP) mechanism at low (or high) temperatures. The EY mechanism also
contributes to the spin relaxation at intermediate
temperature.

We also study the temperature dependence of the spin relaxation. The
behavior also depends on the Mn concentration $x$ as the relevant
spin relaxation mechanisms are different for different $x$. In
the small $x$ regime, the SRT first increases then decreases
with increasing temperature which resembles what was found in $n$-type
quantum wells with low impurity density.\cite{lowT} In the
large $x$ regime, at low temperature the $s$-$d$ exchange scattering
mechanism is dominant, which, however, is independent of the
temperature. The temperature dependence is hence very weak. As the
temperature increases, the EY and BAP mechanisms become more and
more important, which lead to a fast decrease of the SRT with
temperature. In the medium $x$ regime, the DP mechanism is also
important. As the momentum scattering is dominated by the
electron-impurity scattering which changes slowly with temperature,
the increase of the inhomogeneous broadening leads to the decrease
of the SRT. The SRT due to the BAP and EY mechanisms also decreases
with increasing temperature. Consequently, the SRT also decreases monotonically
with increasing temperature in the medium $x$ regime.

We then address the photo-excitation density dependence of the
SRT. The behavior is different for different temperature and $x$. At
low temperature, as the electron system is in the degenerate regime,
the DP mechanism is largely enhanced as the inhomogenous broadening
increases. However, the $s$-$d$ exchange scattering mechanism is
independent of the photo-excitation, and the BAP mechanism changes slowly with
the photo-excitation density as the hole Pauli blocking is very
strong. The EY mechanism is usually less efficient than the DP mechanism.
Consequently, the peak in the $\tau$-$x$ curve moves to larger
$x$ value. The SRTs in the small and medium $x$ regimes decrease with increasing
photo-excitation density as the DP spin relaxation
increases. However, the SRT at large $x$ regime changes little as the
spin relaxation is dominated by the $s$-$d$ exchange scattering
mechanism. The behavior is quite different at high temperature where
the electron system is in the non-degenerate regime. In the small $x$
regime, where the momentum scattering is dominated by the
carrier-carrier Coulomb scattering as the impurity density is low. The
SRT increases with increasing photo-excitation density as the carrier-carrier Coulomb
scattering increases with increasing carrier density. In the medium $x$ regime,
where the electron-impurity scattering is dominant.
As the screening increases with increasing carrier density, the electron-impurity
 scattering decreases with increasing photo-excitation density. Hence the
 SRT due to the DP mechanism decreases with increasing photo-excitation
 density. The EY mechanism is less important than the DP mechanism in
  this regime. Moreover, as
the BAP mechanism becomes important, the SRT decreases with increasing
photo-excitation density as the hole density increases. In the large
$x$ regime, the holes mainly come from the Mn dopants and the spin
relaxation is dominated by the BAP mechanism, the SRT hence changes
little with photo-excitation.

We also discuss the magnetic field dependence of the SRT. We find that
the main effect at low magnetic field is the mixture of the in-plane
and out-of-plane spin relaxations. The spin relaxation due to the BAP
and $s$-$d$ exchange scattering mechanisms is isotropic, whereas that
due to the DP and EY mechanism is anisotropic. For the DP mechanism
the in-plane spin relaxation is slower than the out-of-plane one,
whereas for the EY mechanism, the in-plane one is faster than the
out-of-plane one. Therefore, in small and medium $x$ regimes where the DP
mechanism is more important than the EY mechanism, the magnetic field
induces an abrupt increase of the SRT due to the mixing of the
in-plane and out-of-plane spin relaxations.\cite{Dohrmann} In large $x$
regime, the EY mechanism is more important than the DP mechanism, and
the SRT hence decreases abruptly with increasing magnetic field. Another
important effect of the magnetic field is that it induces an
equilibrium Mn spin polarization and thus enhances the $s$-$d$ exchange
scattering mechanism. Consequently, for large $x$ at low
temperature, where the $s$-$d$ exchange scattering dominates the spin
relaxation, the SRT decreases with increasing magnetic field. In other
regimes, the $s$-$d$ exchange scattering mechanism is unimportant and the
magnetic field dependence of the SRT after the abrupt jump is also
weak. We find that the nonequilibrium spin polarization
transfered from the electron system to the Mn system due to the
$s$-$d$ exchange interaction is much smaller than the electron spin
polarization, which is consistent with the fact that the Mn beats are
not observed in experiments.\cite{Awsch}

\begin{acknowledgments}
  This work was supported by the National Natural Science Foundation of
  China under Grant No.~10725417, the National Basic
  Research Program of China under Grant No.~2006CB922005 and the
  Knowledge Innovation Project of Chinese Academy of Sciences,
the Robert-Bosch Stiftung, as well
  as by the DFG via SFB 689 and SPP 1285.
\end{acknowledgments}


\begin{thebibliography}{0}
\bibitem{Munekata1} H. Munekata, H. Ohno, S. von Moln\'ar,
  A. Segm\"uller, L. L. Chang, and L. Esaki, Phys. Rev. Lett.
  {\bf 63}, 1849 (1989).
\bibitem{Ohno1} H. Ohno, H. Munekata, T. Penney, S. von Moln\'ar, and
  L. L. Chang, Phys. Rev. Lett. {\bf 68}, 2664 (1992).
\bibitem{Munekata2} H. Munekata, A. Zaslavsky, P. Fumagalli, and
  R. J. Gambino, Appl. Phys. Lett. {\bf 63}, 2929 (1993).
\bibitem{Ohno2} H. Ohno, A. Shen, F. Matsukura, A. Oiwa, A. Endo,
  S. Katsumoto, and Y. Iye, Appl. Phys. Lett. {\bf 69}, 363 (1996).
\bibitem{Hayashi} T. Hayashi, M. Tanaka, K. Seto, T. Nishinaga, and
  K. Ando, Appl. Phys. Lett. {\bf 71}, 1825 (1997).
\bibitem{VanEsch} A. Van Esch, L. Van Bockstal, J. De Boeck,
  G. Verbanck, A. S. van Steenbergen, P. J. Wellmann, B. Grietens,
  R. B. F. Herlach, and G. Borghs, Phys. Rev. B {\bf 56}, 13103 (1997).
\bibitem{Ohno3} H. Ohno, Science {\bf 281}, 951 (1998).
\bibitem{FStheory} T. Jungwirth, J. Sinova, J. Mas\v ek, J. Ku\v cera,
  and A. H. MacDonald, Rev. Mod. Phys. {\bf 78}, 809 (2006).
\bibitem{MSreview} {\em Magnetic Semiconductors}, ed. by
  J. K. Furdyna and J. Kossut, Semiconductor and Semimetals Vol. 25
  (Academic, New York, 1988); {\em Diluted Magnetic Semiconductors},
  ed. by M. Balkanski and M. Averous (Plenum, New York, 1991).
\bibitem{spintronics} {\em Semiconductor Spintronics and Quantum
    Computation}, ed. by D. D. Awschalom, D. Loss, and N. Samarth
  (Springer-Verlag, Berlin, 2002);  I. \v Zuti\'c, J. Fabian, and S. Das Sarma,
Rev. Mod. Phys. {\bf 76}, 323 (2004); J. Fabian, A. Matos-Abiague, C. Ertler,
P. Stano, and I. \v Zuti\'c, Acta Phys. Slov. {\bf 57}, 565 (2007);
{\em Spin Physics in Semiconductors}, ed. by
M. I. D'yakonov (Springer, Berlin, 2008), and references therein.
\bibitem{Wolf} S. A. Wolf, D. D. Awschalom, R. A. Buhrman, J.
M. Daughton, S. von Moln\'ar, M. L. Roukes, A. Y. Chtchelkanova, and
D. M. Treger, Science {\bf 294}, 1488 (2001).
\bibitem{Dietl} T. Dietl, in {\em Modern Aspects of Spin Physics}, ed.
  by J. Fabian, vol. 712 (Springer, Berlin, 2007) p. 1-46.
\bibitem{Fabian} C. Ertler and J. Fabian, Phys. Rev. Lett. {\bf 101},
  077202 (2008).
\bibitem{Back} F. Maccherozzi, M. Sperl, G. Panaccione, J. Min\'ar,
  S. Polesya, H. Ebert, U. Wurstbauer, M. Hochstrasser, G. Rossi,
  G. Woltersdorf, W. Wegscheider, and C. H. Back,
  Phys. Rev. Lett. {\bf 101}, 267201 (2008).
\bibitem{Wagner} K. Wagner, D. Neumaier, M. Reinwald, W. Wegscheider,
  and D. Weiss, Phys. Rev. Lett. {\bf 97}, 056803 (2006).
\bibitem{Saha} D. Saha, L. Siddiqui, P. Bhattacharya, S. Datta, D. Basu,
  and M. Holub, Phys. Rev. Lett. {\bf 100}, 196603 (2008).
\bibitem{Ciorga} For recent advancement in spin injection from
  ferromagnetic semiconductors, see, M. Ciorga, A. Einwanger,
  U. Wurstbauer, D. Schuh, W. Wegscheider, and D. Weiss, arXiv:0809.1736.
\bibitem{Ruster} C. R\"uster, T. Borzenko, C. Gould, G. Schmidt,
  L. W. Molenkamp, X. Liu, T. J. Wojtowicz, J. K. Furdyna, Z. G. Yu,
  and M. E. Flatt\'e, Phys. Rev. Lett. {\bf 91}, 216602 (2003).
\bibitem{Tang} H. X. Tang, R. K. Kawakami, D. D. Awschalom, and
  M. L. Roukes, Phys. Rev. Lett. {\bf 90}, 107201 (2003).
\bibitem{Ohno4} H. Ohno, D. Chiba, F. Matsukura, T. Omiya, E. Abe,
  T. Dietl, Y. Ohno, and K. Ohtani, Nature {\bf 408}, 944 (2000).
\bibitem{opt-mag} J. Fern\'andez-Rossier, C. Piermarocchi, P. Chen,
  A. H. MacDonald, and L. J. Sham, Phys. Rev. Lett. {\bf 93}, 127201
  (2004); J. Wang, C. Sun, J. Kono, A. Oiwa, H. Munekata,
  L. Cywinski, and L. J. Sham, Phys. Rev. Lett. {\bf 95}, 167401 (2005);
  J. Wang, I. Cotoros, K. M. Dani, X. Liu, J. K. Furdyna, and
  D. S. Chemla, Phys. Rev. Lett. {\bf 98}, 217401 (2007);
  Y. Hashimoto, S. Kobayashi, and H. Munekata, Phys. Rev. Lett. {\bf
  100}, 067202 (2008); J. Wang, I. Cotoros, X. Liu, J. Chovan,
  J. K. Furdyna, I. E. Perakis, and D. S. Chemla, arXiv:0804.3456.
\bibitem{Awsch} M. Poggio, R. C. Myers, N. P. Stern, A. C. Gossard,
  and D. D. Awschalom, Phys. Rev. B {\bf 72}, 235313 (2005).
\bibitem{Schulz} R. Schulz, T. Korn, D. Stich, U. Wurstbauser,
  D. Schuh, W. Wegscheider, and C. Sch\"uller, Physica E {\bf 40},
  2163 (2008).
\bibitem{Korn} T. Korn, R. Schulz, S. Fehringer, U. Wurstbauer, D. Schuh,
  W. Wegscheider, M. W. Wu, and C. Sch\"uller, arXiv:0809.3654.
\bibitem{MOreview} K. S. Burch, D. D. Awschalom, D. N. Basov,
  J. Magn. Magn. Mater. {\bf 320}, 3207 (2008).
\bibitem{opt-or} F. Meier and B. P. Zakharchenya, {\em Optical
    Orientation} (North-Holland, Amsterdam, 1984).
\bibitem{DP} M. I. D'yakonov and V. I. Perel',
  Zh. \'Eksp. Teor. Fiz. {\bf 60}, 1954 (1971) [Sov. Phys. JETP {\bf
  33}, 1053 (1971)]; Fiz. Tverd. Tela (Leningrad) {\bf 13}, 3581
  (1971) [Sov. Phys. Solid State {\bf 13}, 3023 (1972)].
\bibitem{BAP} G. L. Bir, A. G. Aronov, and G. E. Pikus,
  Zh. \'Eksp. Teor. Fiz. {\bf 69}, 1382 (1975) [Sov. Phys. JETP {\bf
  42}, 705 (1976)].
\bibitem{EY} Y. Yafet, Phys. Rev. {\bf 85}, 478 (1952); R. J. Elliot,
  Phys. Rev. {\bf 96}, 266 (1954).
\bibitem{lowT} J. Zhou, J. L. Cheng, and M. W. Wu, Phys. Rev. B {\bf
    75}, 045305 (2007).
\bibitem{wu-bap} J. Zhou and M. W. Wu, Phys. Rev. B {\bf 77}, 075318
  (2008).
\bibitem{bulk} J. H. Jiang and M. W. Wu, arXiv:0812.0862.
\bibitem{Edmonds} K. W. Edmonds, P. Bogus\l awski, K. Y. Wang,
  R. P. Campion, S. N. Novikov, N. R. S. Farley, B. L. Gallagher,
  C. T. Foxon, M. Sawicki, T. Dietl, M. Buongiorno Nardelli,
  and J. Bernholc, Phys. Rev. Lett. {\bf 92}, 037201 (2004).
\bibitem{wu-early} M. W. Wu and C. Z. Ning, Eur. Phys. J. B {\bf 18},
  373 (2000); M. W. Wu and H. Metiu, Phys. Rev. B {\bf 61}, 2945
  (2000); M. W. Wu, J. Phys. Soc. Jpn. {\bf 70}, 2195 (2001).
\bibitem{highP} M. Q. Weng and M. W. Wu, Phys. Rev. B {\bf 68}, 075312 (2003).
\bibitem{hot-e} M. Q. Weng, M. W. Wu, and L. Jiang, Phys. Rev. B
  {\bf 69}, 245320 (2004).
\bibitem{multi-band} M. Q. Weng and M. W. Wu, Phys. Rev. B {\bf 70}, 195318
  (2004).
\bibitem{wu-hole} C. L\"u, J. L. Cheng, and M. W. Wu, Phys. Rev. B {\bf 73}, 125314
  (2006).
\bibitem{multi-valley} P. Zhang, J. Zhou, and M. W. Wu, Phys. Rev. B
  {\bf 77}, 235323 (2008).
\bibitem{terahertz} J. H. Jiang, M. W. Wu, and Y. Zhou, Phys. Rev. B
  {\bf 78}, 125309 (2008).
\bibitem{Ji} X. Z. Ruan, H. H. Luo, Y. Ji, Z. Y. Xu, and V. Umansky,
  Phys. Rev. B {\bf 77}, 193307 (2008).
\bibitem{Lai} L. H. Teng, P. Zhang, T. S. Lai, and M. W. Wu,
  Europhys. Lett. {\bf 84}, 27006 (2008).
\bibitem{wu-exp-hP} D. Stich, J. Zhou, T. Korn, R. Schulz, D. Schuh,
  W. Wegscheider, M. W. Wu, and C. Sch\"uller, Phys. Rev. Lett. {\bf
    98}, 176401 (2007); Phys. Rev. B {\bf 76}, 205301
  (2007).
\bibitem{tobias} T. Korn, D. Stich, R. Schulz, D. Schuh,
  W. Wegscheider, and C. Sch\"uller, arXiv:0811.0720.
\bibitem{wu-strain} L. Jiang and M. W. Wu, Phys. Rev. B {\bf 72},
  033311 (2005).
\bibitem{wu-exp} D. Stich, J. H. Jiang, T. Korn, R. Schulz, D. Schuh,
  W. Wegscheider, M. W. Wu, and C. Sch\"uller, Phys. Rev. B {\bf 76},
  073309 (2007).
\bibitem{Awsch-exp} A. W. Holleitner, V. Sih, R. C. Myers, A. C. Gossard,
  and D. D. Awschalom, New J. Phys. {\bf 9}, 342 (2007).
\bibitem{Zheng} F. Zhang, H. Z. Zheng, Y. Ji, J. Liu, and G. R. Li,
  Europhys. Lett. {\bf 83}, 47006 (2008)
\bibitem{Zheng2} F. Zhang, H. Z. Zheng, Y. Ji, J. Liu, and G. R. Li,
  Europhys. Lett. {\bf 83}, 47007 (2008).
\bibitem{Madelung} E. T. Yu, J. O. McCaldin, and
  T. C. McGill, Solid State Phys. {\bf 46}, 1 (1992).
\bibitem{Haug} H. Haug and A.-P. Jauho, {\em Quantum Kinetics in
  Transport and Optics of Semiconductors} (Springer, Berlin, 1998).
\bibitem{Dresselhaus} G. Dresselhaus, Phys. Rev. {\bf 100}, 580
  (1955).
\bibitem{Rashba} Y. A. Bychkov and E. I. Rashba, J. Phys. C
  {\bf 17}, 6039 (1984); JETP Lett. {\bf 39}, 78 (1984).
\bibitem{Flatte} W. H. Lau and M. E. Flatt\'e, Phys. Rev. B {\bf 72},
  161311 (2005).
\bibitem{Myers} R. C. Myers, M. H. Mikkelsen, J.-M. Tang,
  A. C. Gossard, M. E. Flatt\'e, and D. D. Awschalom, Nature Materials
  {\bf 7}, 203 (2008). 
\bibitem{para} {\em Semiconductors}, Landolt-B\"{o}rnstein, New
  Series, Vol.\ 17a, ed. by O. Madelung (Springer, Berlin, 1987).
\bibitem{Ekardt} W. Ekardt, K. L\"osch, and D. Bimberg,
  Phys. Rev. B {\bf 20}, 3303 (1979).
\bibitem{Maialle} M. Z. Maialle, Phys. Rev. B {\bf 54}, 1967 (1996).
\bibitem{Sanvito} S. Sanvito, P. Ordej\'on, and N. A. Hill,
  Phys. Rev. B {\bf 63}, 165206 (2001).
\bibitem{Harley} W. J. Leyland, G. H. John, R. T. Harley,
  M. M. Glazov, E. L. Ivchenko, D. A. Ritchie, I. Farrer,
  A. J. Shields, and M. Henini, Phys. Rev. B {\bf 75}, 165309 (2007).
\bibitem{Vignale} G. F. Giulianni and G. Vignale, {\em Quantum Theory of the
  Electron Liquid} (Cambridge University Press, Cambridge, 2005).
\bibitem{localize} M. Syperek, D. R. Yakovlev, A. Greilich,
  J. Misiewicz, M. Bayer, D. Reuter, and A. D. Wieck,
  Phys. Rev. Lett. {\bf 99}, 187401 (2007).
\bibitem{Astakhov} G. V. Astakhov, R. I. Dzhioev, K. V. Kavokin,
  V. L. Korenev, M. V. Lazarev, M. N. Tkachuk, Yu. G. Kusrayev,
  T. Kiessling, W. Ossau, and L. W. Molenkamp, Phys. Rev. Lett. {\bf
  101}, 076602 (2008).
\bibitem{Sliwa} C. \'Sliwa and T. Dietl, Phys. Rev. B {\bf 78}, 165205
  (2008).
\bibitem{Dohrmann} S. D\"ohrmann, D. H\"agele, J. Rudolph,
  M. Bichler, D. Schuh, and M. Oestreich, Phys. Rev. Lett. {\bf 93},
  147405 (2004).
\bibitem{Linder} N. Linder and L. J. Sham, Physica E {\bf 2},
  412 (1998).
\bibitem{Crooker} S. A. Crooker, D. D. Awschalom, J. J. Baumberg,
  F. Flack, and N. Samarth, Phys. Rev. B {\bf 56}, 7574 (1997).

\end{thebibliography}
\end{document}